\documentclass[fontsize=11pt,a4paper,toc=flat]{arxiv-article}
\usepackage{Definitions}
\acused{bec}
\acused{bcs}

\DeclareDocumentCommand\Op{ m}{
  \mathinner{\mathcal{O}_{#1}}
}

\usepackage{pgfplots}
\pgfplotsset{compat=1.18} % Recommended for newer versions
\usetikzlibrary{arrows.meta} 

\newsavebox{\lowenergyplot}
\newsavebox{\linearplot}
\newsavebox{\highenergyplot}
\newsavebox{\highenergyarrow}
\newsavebox{\midenergyarrow}
\newsavebox{\lowenergyarrow}

\sbox{\lowenergyplot}{%
    \begin{tikzpicture}
        \begin{axis}[
            xmin=1.4, xmax=1.8,
            ymin=0, ymax=.8,
            xlabel={}, %
            ylabel={}, %
            xtick=\empty, %
            ytick=\empty, %
            grid=none, %
            axis lines=middle, %
            tick label style={font=\empty}, %
            scale only axis,
            scale=.1
        ]
        \addplot[
            domain=1:2,
            samples=100,
            smooth,
            blue,
            thick
        ] {x - 2/x};

        \end{axis}
    \end{tikzpicture}%
} %

\sbox{\linearplot}{%
    \begin{tikzpicture}
        \begin{axis}[
            xmin=0, xmax=1,
            ymin=0, ymax=1,
            xlabel={}, %
            ylabel={}, %
            xtick=\empty, %
            ytick=\empty, %
            grid=none, %
            axis lines=middle, %
            tick label style={font=\empty}, %
            scale only axis,
            scale=.1
        ]
        \addplot[
            domain=.2:1,
            samples=100,
            smooth,
            blue,
            thick
        ] {x};

        \end{axis}
    \end{tikzpicture}%
} %

\sbox{\highenergyplot}{%
    \begin{tikzpicture}
        \begin{axis}[
            xmin=0, xmax=1,
            ymin=0, ymax=1,
            xlabel={}, %
            ylabel={}, %
            xtick=\empty, %
            ytick=\empty, %
            grid=none, %
            axis lines=middle, %
            tick label style={font=\empty}, %
            scale only axis,
            scale=.1
        ]
        \addplot[
            domain=.4:1,
            samples=100,
            smooth,
            blue,
            thick
        ] {x^2};

        \end{axis}
    \end{tikzpicture}%
} %

\sbox{\highenergyarrow}{
  \begin{tikzpicture}[scale=.15]
  \shade[left color=red, right color=blue] (0,0) rectangle (10,1);
  \draw[-{Triangle[width=18pt, length=10pt]}, line width=10pt, blue] (9,4) -- (9,1) ;
\end{tikzpicture}
}

\sbox{\midenergyarrow}{
  \begin{tikzpicture}[scale=.15]
  \shade[left color=red, right color=blue] (0,0) rectangle (10,1);
  \draw[-{Triangle[width=18pt, length=10pt]}, line width=10pt, blue!50!red] (5,4) -- (5,1) ;
\end{tikzpicture}}

\sbox{\lowenergyarrow}{
  \begin{tikzpicture}[scale=.15]
  \shade[left color=red, right color=blue] (0,0) rectangle (10,1);
  \draw[-{Triangle[width=18pt, length=10pt]}, line width=10pt, red] (1,4) -- (1,1) ;
\end{tikzpicture}
}

\usepackage{marginnote}
\usepackage{hyperref}
\graphicspath{{./plots/}{./FIGS/}}

\preprint{\texttt{NT@UW-25-16}}

\newcommand{\OfficialTitle}{
  Trapping-potential dependence of
  the unitary Fermi gas at the BCS-BEC crossover
}
\title{\setstretch{1.4}
	{\color{Thoughtless}\textls[-20]{\OfficialTitle}}
}

\hypersetup{pdfauthor={Silas R.~Beane, Domenico Orlando, Susanne Reffert},pdftitle={\OfficialTitle},%
	colorlinks=true,linkcolor=ThoughtYouWere,citecolor=ThoughtYouWere,urlcolor=ThoughtYouWere}

\author{%
	\begin{minipage}{.94\textwidth}
		\begin{center} \dosserif%
			{\small
           \textbf{Silas~R.~Beane}\textsuperscript{\ding{74}},
           \textbf{Adèle~Le~Borgne}\textsuperscript{\ding{73}},
				\textbf{Domenico~Orlando}\textsuperscript{\ding{72}\ding{73}}, and
  				\textbf{Susanne~Reffert}\textsuperscript{\ding{73}} 
			}
		\end{center}
  		\authorBlock{\ding{74}}{\dosserif{}%
                        Department of Physics,\\
                        University of Washington,\\
                        Seattle, WA 98195}
		\authorBlock{\ding{73}}{\dosserif{}%
			Albert Einstein Center for Fundamental Physics,\\
			Institute for Theoretical Physics, University of Bern,\\
			Sidlerstrasse 5, CH-3012 Bern, Switzerland}
		\authorBlock{\ding{72}}{\dosserif{}%
			INFN sezione di Torino.\\
			via Pietro Giuria 1, 10125 Torino, Italy}
	\end{minipage}
}

\date{}

\begin{document}

\numberwithin{equation}{section}

\begin{titlepage}

	%\newgeometry{top=23.1mm,bottom=46.1mm,left=34.6mm,right=34.6mm}

\maketitle

	\thispagestyle{empty}

   \vfill
   %\dosserif{}

   \abstract{ \normalfont{}\noindent{} Cold-atom experiments which
     measure Fermi-gas properties near unitarity confine fermionic
     atoms to a region of space using trapping potentials of various
     shapes. The presence of a trapping potential introduces a new
     characteristic physical scale in the superfluid \acs{eft} description of the unitary Fermi gas which,
     \textit{inter alia}, describes the acoustic branch of excitations in
     the far infrared well below the scale of the superfluid gap. In
     this \acs{eft} there is a clear hierarchy of scales, and
     corrections to the homogeneous system due to the trapping
     potential may be organized into three regions with distinct power
     counting that relies on both the \acs{eft} derivative expansion,
     and the \acs{wkb} approximation, which is an expansion in
     gradients of the trapping potential. The energy spectrum of the
     superfluid system is obtained in each of the regions by explicit
     computation of the phonon-field fluctuations, and by the
     modifications to the dynamic structure factor due to the
     corresponding density fluctuations. This work presents a systematic and quantitative method for treating the presence of a trapping potential, which is essential for interpreting experimental realizations of the unitary Fermi gas. It provides clear predictions for how the trapping potential modifies the dispersion relation's curvature, a key factor in characterizing the relaxation mechanisms of the superfluid.  The most significant
     deviations from linear dispersion due to the trapping potential
     are found in the far infrared region of the superfluid \acs{eft}. }
          %\kant[1]}
	\vfill

\end{titlepage}

%\restoregeometry{}

\setstretch{1.1}
%%%%%%%%%%%%%%%%%%%%%%%%%%%%%
\tableofcontents

%\listoffigures
%\newpage
% * introduction
%%%%%%%%%%%%%%%%%%%%%%%%%%%%%%%%%%%%%%%%%%%%%%%%%%%%%
\section{Introduction}
\label{sec:intro}

\noindent A gas of spin-$1/2$ fermionic atoms, whose effective
two-body interactions may be tuned using Feshbach resonances, at and
near unitarity, provides a remarkable strongly-interacting many-body
system that can be studied experimentally to a high degree of accuracy
using cold-atom techniques, while also being tractable theoretically
due to the large degree of symmetry and clear scale
separation~\cite{Zwerger2012}. From the perspective of the fundamental
field theoretic formulation of the strongly-interacting fermion
system, the absence of a small expansion parameter implies the
necessity of nonperturbative methods for a first principles
computation of observables. So, for instance, the energy density of
the system, characterized by the so-called Bertsch parameter, is
computed using \ac{qmc} simulations~\cite{Carlson_2011} and measured in
cold-atom experiments~\cite{Ku_2012,Z_rn_2013}.  However, at
unitarity, and indeed over the entire \acs{bcs}--\acs{bec} crossover
region, where the large scattering length approaches infinity
(unitarity) from positive (\acs{bec}) and negative (\acs{bcs}) values,
the infrared degrees of freedom are superfluid and are therefore
described by a single Goldstone boson (phonon) degree of
freedom~\cite{Greiter:1989qb,Son:2005rv,Ma_es_2009}. At unitarity the
\ac{eft} of the Goldstone mode is Schr\"odinger invariant and
therefore is a \ac{nrcft} which admits powerful techniques for
computing observables such as the state-operator
correspondence~\cite{Nishida:2007pj} and the large-charge
expansion~\cite{Hellerman:2015nra,Gaume:2020bmp,Favrod:2018xov,Kravec:2018qnu,Orlando:2020idm,Pellizzani:2021hzx,Hellerman:2023myh,Hellerman:2020eff,Beane:2024kld,Beane:2025tum}, where an \ac{eft} can be formulated as an expansion in powers of $1/Q$, where $Q$ is the particle number. This is the approach that will be adopted in the following.
The conformal data which is encoded in the low-energy constants of the
\ac{eft}---including the Bertsch parameter---is external to the
\ac{eft} and must be obtained via theoretical modeling, experimental
measurement, and/or numerical simulation.  Promising systematic
methods for obtaining this data include the so-called
$\epsilon$-expansion~\cite{Nishida:2006br}, and the large-$N$
expansion~\cite{Hellerman:2023myh}, where $N$ is the number of fermion
flavors. At leading order in the $1/N$ expansion, this model
reproduces well-known mean-field theory results, but has the advantage
of providing arbitrary systematic improvements in a perturbative
expansion in powers of $1/N$.

A main focus of this paper is on the phonon dispersion relation which 
describes the acoustic excitation branch of the superfluid throughout
the \acs{bcs}--\acs{bec} crossover. In the absence of a confining potential, the dispersion relation may be written as~\cite{castin2025},
\begin{equation}
  \label{eq:general-dispersion}
  q_0(q) = c_s q \pqty*{1 + \frac{\gamma}{8 c_s^2}q^2 + \mathcal{O}(q^4 \log q)}.
\end{equation}
Here $q_0$ is the energy, $q$ is the momentum transfer, $c_s$ is the
speed of sound, and $\gamma$ is a parameter which governs the
deviation of the spectrum from linearity. The form of the dispersion
relation is a simple consequence of the spontaneously broken particle
number in the superfluid phase~\cite{Son:2005rv}. Throughout we work
in units where $\hbar=M=1$, where $M$ is the atomic mass . At unitarity, the \ac{eft} is a \ac{nrcft}, constrained by
Schr\"odinger invariance, with the sole scale given by the chemical
potential, $\mu$, which is generated by spontaneous symmetry
breaking. Hence, the expansion of the dispersion relation given in
Eq.~(\ref{eq:general-dispersion}) is in $q$ relative to the \ac{uv}
scale $\mu$. Observables of special interest in the superfluid
\ac{eft} at unitarity are correlation functions at fixed
particle-number charge $Q$. In this paper the focus is on the
fluctuations of the Goldstone boson field relative to the scale $\mu$
of the fermion chemical potential which is fixed by the number of particles which plays the role of a
$U(1)$ charge. Working at large charge directly translates to working
at large $\mu$\footnote{Operationally this corresponds to a change of
  ensemble from Canonical (particle number specified) to Grand
  Canonical (chemical potential specified).}.

The sign of $\gamma$ is essential for understanding the gas relaxation
mechanism at low temperature. So, for instance, if $\gamma>0$ (a
convex branch) then the three-phonon Beliaev--Landau process is
kinematically allowed, whereas if $\gamma<0$ (a concave branch) then
phonon damping can occur only at $T\neq 0$ via the as-yet unobserved
four-phonon Landau--Khalatnikov process~\cite{Kurkjian_2017}.  A recent
experiment using Bragg spectroscopy~\cite{PhysRevLett.128.100401} has
found $\gamma_{\scriptstyle EXP}=-0.085(8)$ (a concave
branch). However, this result has been questioned in
Refs.~\cite{Castin_2024,castin2025} based on potentially uncontrolled
finite-temperature effects and a fitting-range bias which favors
higher momentum transfers.

Atomic physics experiments which explore the \acs{bcs}--\acs{bec}
crossover suffer from nontrivial experimental influences which must be
carefully considered in extracting the dispersion-relation
parameters. In particular, the fermionic atoms are contained
in a trapping potential at a finite temperature.  The impact of these
environmental scales on the many-body system must be understood
quantitatively in order to correctly interpret the implications of the
experimental data for the zero-temperature homogeneous system. This
paper will focus on developing a systematic \ac{eft} analysis of the effect
of the trapping potential on the phonon fluctuations, at zero temperature.

The presence of a trapping potential, here denoted $V$, breaks the
translational symmetry of the spatially uniform system, and introduces
a new scale in the superfluid \ac{eft} which can be taken as the size
of the confined ``droplet'' or ``cloud'' of superfluid matter, defined as the
distance scale at which the leading order of the density vanishes. The \ac{eft} hierarchy
may then be set up to probe distance scales that are larger than the
typical inter-particle separation fixed by the density, but less than
the size of the confined droplet. A standard tool for dealing with
spatially dependent densities is the \ac{lda}~\cite{HoZhou2010}, which
amounts to shifting the fermion chemical potential by $V$ so that it
becomes spatially dependent. Here we use the superfluid \ac{eft} to
investigate, systematically, corrections to the \ac{lda} which
account for the effect of the trapping potential on the dispersion
relation which describes the acoustic branch of excitations in the
superfluid. In practice, this amounts to solving the phonon equation
of motion in the presence of fluctuations on top of the homogeneous
ground state. This turns out be achievable for an arbitrary (static)
potential, $V$, using the \ac{wkb} approximation together with the
homogeneous \ac{eft} power counting.

As the potential $V$ breaks spatial translations, the three momentum components are no longer good
quantum numbers in the trap.  Strictly speaking, the
dispersion relation which relates energy to momentum is not defined.
A natural way to connect to atomic physics measurements is to instead
consider the dynamic structure factor $S$, which is naturally defined
as a matrix element of local operators in the superfluid \ac{eft}.  In
translation-invariant systems, $S$ is a delta function with support
on the curve of the dispersion relation. We can define an approximate
dispersion relation as the curve in the energy-momentum plane along
which the dynamic structure factor is peaked. The flatter the
potential, the better the dispersion relation is approximated.

Unlike the dispersion relation, the energy spectrum of the fluctuation
is well-defined because of time-translation invariance. Due to the
compactness of the system, the spectrum is discrete as will be proven
below (see Section~\ref{sec:fluctuation-spectrum}). In general, the
spectrum and the dispersion relation carry distinct information. They
have the same functional form only in translation-invariant systems.

\bigskip

For purposes of clarity, it is worth giving an overview here of some
of the main points that should be kept in mind while reading this
paper.  Our goal in what follows is to describe the small density
fluctuations \(\delta\rho\) of the fermion cloud.  First, consider the
symmetries.  The trap is time independent and spherically symmetric so it only depends on the distance from the center. The system preserves time-translation
invariance and rotational invariance.
This implies that \(\delta \rho\) can be decomposed into modes of the form
  \begin{align}
    \delta \rho_{nlm} &= e^{i q_0(n,\ell) t} f_{n \ell}(r) Y_{\ell m}(\theta, \phi) \, , & \ell = 0, 1, \dots \, , m = -\ell, ..., \ell \, ,
  \end{align}
where \(Y_{\ell m}\) are spherical harmonics.  The function \(q_0 =
q_0(n, \ell)\) identifies the allowed values for the energy of the
fluctuations: it is the \emph{spectrum}.  If the potential is
confining, then the spectrum is discrete and the parameter \(n\) is
an integer. The information contained in the radial function \(f_{n \ell}(r)\)
can be extracted by defining the response function 
  \begin{equation}
    \chi( t - t';\mathbf{r}, \mathbf{r}') = \ev{ \delta \rho(t,\mathbf{r}) \delta \rho(t',\mathbf{r}')}_T \, ,
  \end{equation}
where \(\ev{\dots}_T\) represents time ordering.
  
In the special case of invariance under translations in space, the radial function is
an exponential (or a sum of exponentials as needed to satisfy the boundary
conditions) and the Fourier transform from \(\mathbf{r}\) to
\(\mathbf{q}\) is proportional to a delta function that fixes a linear
relation between \(q\) and \(n\).  For example, for a spherical box of radius \(R_{cl}\), with zero potential inside the box and infinite potential outside
  \begin{equation}
    q(n) = \frac{\pi n}{R_{cl}} \, .
  \end{equation}
When the dynamic structure factor, which is related to the absorptive
part of the response function, is a sum of delta functions, the
spectrum equation \(q_0 = q_0(n)\) has the same functional form as the
dispersion relation \(q_0 = q_0(q)\), and one is justified in thinking
of \(n\) as a discrete momentum.

By contrast, with broken spatial translations, the spectrum and dispersion relation
contain distinct information.  However, if we compute the
fluctuations in a \ac{wkb} expansion which is equivalent to a gradient expansion in the potential, at leading order, \(f_{nl}(r)\)
will take the same form as in the case of no potential, but with the
substitution of the fermion chemical potential, \(\mu \to \mu - V(r)\).  This is nothing other than the
\ac{lda}.  The Fourier integral will not give a delta function, but
some appropriately peaked smooth function.  And the relationship
between spectrum and dispersion relation remains true only at \ac{lo}.

\bigskip

Our main findings on the dependence of the spectrum and dispersion relation on the trapping potential are:
\begin{enumerate}
\item In first approximation, the behavior of the system depends on the details of the trapping potential close to the center of the trap.
\item The spectrum is more dense for steep potentials ($k$ smaller for $V(r) \sim r^{2k}$, $k\ne 0$) for fixed size of the cloud \(R_{cl}\).
  The flatter the potential in the center ($k$ larger), the more the special case of a flat spherical box potential is approximated and the states become equidistant.
  Asymptotically, for $k\rightarrow \infty$, we reproduce the spectrum of the spherical box \(q_0 = \sqrt{2\mu/3} \pi n/ R_{cl}\).
\item The dynamic structure factor is more peaked for flatter potentials.
  In turn this means that the dispersion relation is more precise.
\item At low energies (\(q_0 \lessapprox \order{\sqrt{\mu}/R_{cl}}\)), the dispersion relation receives corrections that can be concave.
  The leading effect depends on the details of the trap and is suppressed for flatter potentials.
\item At high energies (\(q_0 \gtrapprox \order{\sqrt{\mu}/R_{cl}} \)), the dispersion relation receives a correction that depends on the low-energy parameters in the \ac{eft} and not on the details of the trap.
  Given the present knowledge of these parameters (that cannot be computed within the \ac{eft} framework), this correction is always convex.
\end{enumerate}

\bigskip

This paper is organized as follows. Section~\ref{sec:sfeft} introduces
the superfluid \ac{eft} and reviews what is currently known about the
low-energy constants that enter up to \ac{nlo} in the homogeneous
system. A scale hierarchy in the presence of the trapping potential is
constructed in Section~\ref{sec:scale-separation}. The fluctuation
spectrum and the dynamic structure factor are investigated in the
various tractable regions in Section~\ref{sec:fluctuation-spectrum}
and Section~\ref{sec:dynamic-structure-factor},
respectively. Conclusions are given in
Section~\ref{sec:conclusions}. Finally, in an Appendix we provide a
derivation of the dynamic structure factor and response function in
the (second-quantized) path-integral formalism.

% * superfluid EFT
%%%%%%%%%%%%%%%%%%%%%%%%%%%%%%%%%%%%%%%%%%%%%%%%%%%%%
\section{Superfluid effective field theory}%
\label{sec:sfeft}

\paragraph{The unitarity limit.}

Our starting point is the \ac{eft} that describes the low-energy
dynamics of a superfluid at unitarity. As originally observed
in~\cite{Greiter:1989qb,Son:2005rv,Ma_es_2009} the \ac{ir} \ac{dof} is
a Goldstone field $\theta$ controlled by the action~\footnote{Note that
  the form of the next-to-leading (NLO) corrections in
  Eq.~(\ref{eq:eft-lag}) differs from that given in
  Ref.~\cite{Son:2005rv}. The correct---conformally invariant---operator structure is given in
  Ref.~\cite{Ma_es_2009}, which noted that Ref.~\cite{Son:2005rv} made
  an invalid use of the leading-order equations of motion to remove an
  operator which is necessary to maintain conformal invariance at
  \ac{nlo}.}
\begin{equation}\label{eq:eft-lag}
   \Lag = c_0 X^{5/2} + c_1 X^{-1/2}(\nabla X)^2 + c_2X^{1/2} \big\lbrack{(\Delta\theta)^2-3(\nabla \otimes \nabla \theta)^2}\big\rbrack + \ldots,
\end{equation}
where the dots are operators with more derivatives, and 
\begin{equation}
	X =\dot\theta -\tfrac{1}{2} (\nabla \theta)^2 - V \, ,
\end{equation}
with $\dot\theta\equiv \del_t \theta$, is a Galilean-invariant combination that frequently appears as a building block in non-relativistic theories in which the external potential $V$ also appears.
It is convenient to define the shifted field,
\begin{equation}
	\theta(t,\mathbf{r}) = \mu t + \pi(t,\mathbf{r}) \ ,
\end{equation}
where $ \mu t$ is the time-dependent \ac{vev} that appears in the classical ground-state and $\pi(t,\mathbf{r})$ is the (phonon) fluctuation.
The particle-number density, $\rho$, is the variable conjugate to the Goldstone field and is given by
\begin{equation}
	\rho(t,\mathbf{r}) = \frac{\partial\Lag}{\partial\dot\theta} = \frac{\partial\Lag}{\partial X}\ .
\end{equation}

\begin{figure}
  \centering
  \includegraphics[width=.75\textwidth]{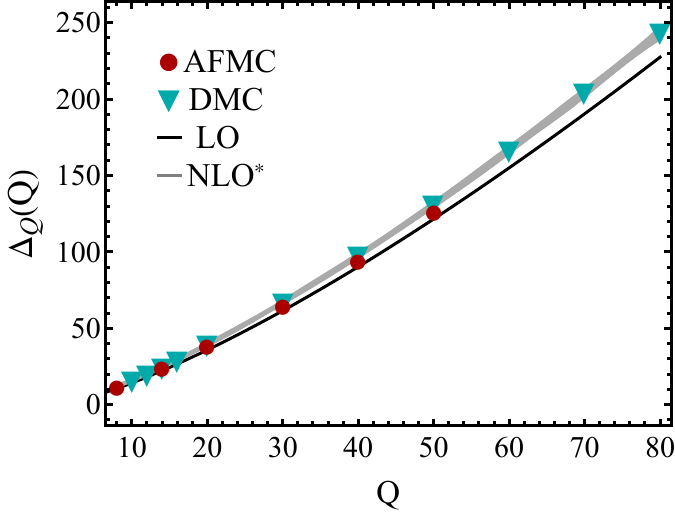}
 \caption{Fit of DMC simulation data~\protect\cite{Carlson:2014pxa} over the range $Q=8-80$, as described in the
  text. The solid black line is leading order in the large-charge
  expansion, and the gray band is the \ac{nlo} fit, including the Casimir
  correction. The simulation data are as described in the text.(Note that the raw data have been provided by the authors of Ref.~\cite{Carlson:2014pxa}.)}
  \label{fig:energy-in-a-trap}
\end{figure}

The $c_i$ are dimensionless low-energy constants that are inputs to the \ac{eft}. Note that
\begin{eqnarray}
  c_0  \ =\ \frac{2^{5/2}}{15\pi^2 \xi^{3/2}} \ ,
  \label{eq:c0}
\end{eqnarray}
with $\xi$ the Bertsch parameter whose value is
available from experiment, $\xi=0.376(4)$
($c_0=0.168(3)$)~\cite{Ku_2012} and from Monte Carlo simulations
$\xi=0.372(5)$ ($c_0=0.166(3)$)~\cite{Carlson_2011}. As the static
transverse response function depends solely on $c_2$, there is a
strict negativity bound $c_2<0$~\cite{Son:2005rv}. There is no direct
knowledge of $c_2$ from experiment or simulation. By contrast, the
parameter $c_1$ is directly related to the static longitudinal
response function, and the energy of $Q$ fermions in a trap, which is
proportional to the large-charge conformal
dimension~\cite{Son:2005rv,Favrod:2018xov,Kravec:2019djc,Hellerman:2021qzz,Beane:2024kld}
\begin{eqnarray}
\Delta_Q(Q) \; =\;  \frac{3^{4/3}}{4} \xi^{1/2}Q^{4/3}\;-\; 3^{2/3 }\sqrt{2}\pi^2 \xi\,c_1\, Q^{2/3} \; +\; \order{Q^{5/9}} \; +\; \ldots \; +\; \frac{1}{3\sqrt{3}}\log Q \ ,
\label{eq:icd}
\end{eqnarray}
where the universal Casimir correction, corresponding to the $Q^0\log(Q)$ term, has been included~\cite{Hellerman:2021qzz}. There have been studies of the energy
of $Q$ fermions in a trap using many methods~\cite{Chang:2007zzd,Endres:2011er,Carlson:2014pxa,Yin_2015}, and one may fit both
$\xi$ and $c_1$ given sufficiently precise data.
Fig.~\ref{fig:energy-in-a-trap} shows simulation data over a range $Q=8-80$ from diffusion \textsc{mc} (\textsc{dmc})
and auxiliary-field \textsc{mc} (\textsc{afmc})~\cite{Carlson:2014pxa}. Here, fitting to the \textsc{dmc} data of Ref.~\cite{Carlson:2014pxa} over the
range  $Q=10-80$, yields $c_1=-0.011(1)$ and $\xi=0.416(7)$ ($c_0=0.143(3)$).

There are various mean-field related determinations of the low-energy
constants\footnote{Leading order in the large-\(N\) expansion
  reproduces mean-field theory.}, all of which are in
agreement~\cite{Ma_es_2009,Kurkjian_2016,Hellerman:2023myh} and find
$\xi=0.5906$ ($c_0=0.0842$), $c_1=-0.0153$, and $c_2=-0.0035$.  In the
$\epsilon$ expansion~\cite{Rupak_2009}, $c_1=-0.0209$,
$c_2=\mathcal{O}(\epsilon^2)$~\footnote{Note that $\epsilon$-expansion conclusions regarding $c_2$,
  associated with the new operator in Eq.~(\ref{eq:eft-lag}), are unchanged.}. It is noteworthy that the mean-field value of $c_1$ is in excellent agreement with the
Monte Carlo simulation data, while the Bertsch parameter is significantly overestimated by the mean-field value.\footnote{A similar observation is made in the relativistic analog of this problem, the large-charge effective theory for the $O(2)$ model. While the low-energy coefficient of the leading operator in the \ac{eft} computed at large $N$ has an error of order $20\%$ with respect to the lattice result, the large-N result for the \ac{nlo} correction in the \ac{eft} lies within the error bars of the lattice computation~\cite{Banerjee:2017fcx,Alvarez-Gaume:2019biu,Dondi:2024vua}.}

The dispersion relation parameters in Eq.~\eqref{eq:general-dispersion} are found to be~\cite{Son:2005rv,Favrod_2018}
\begin{align}%
  \label{eq:EFT-dispersion-parameters1}
  c_s &= \sqrt{\frac{2\mu}{3}} =\sqrt{\frac{\xi}{3}}k_F, & \gamma & = \frac{64}{45} \frac{3 c_2 - c_1}{c_0} \;,
\end{align}
where $k_F$ is the Fermi momentum. It is important to emphasize that at unitarity, the speed of sound is
determined by symmetry.

% * scale separation
%%%%%%%%%%%%%%%%%%%%%%%%%%%%%%%%%%%%%%%%%%%%%%%%%%%%%
\section{Scale separation and perturbative expansions}%
\label{sec:scale-separation}

\paragraph{Strategy.}
In the homogeneous system, the \ac{eft} power counting is simple: all
observable quantities are systematically computable for momentum
transfers well below the scale of the superfluid gap, which, at
unitarity, is set by the chemical potential $\mu$ (times a
dimensionless number of order one)~\cite{Son:2005rv}. The presence of
the trapping potential introduces a new scale---roughly the size of
the fermion cloud---which in turn implies three distinct regions of
interest with the homogeneous power counting enhanced by the \ac{wkb}
expansion.

\paragraph{The small parameter \(\epsilon\).}
To identify the controlling parameters it is convenient to formulate dimensionless variables.
A natural length scale is provided by the classical radius of the particle cloud, \emph{i.e.} the distance from the origin at which the charge density vanishes in the \ac{lo} \ac{eft},
\begin{equation}
  \mu - V(R_{cl}) = 0 .
\end{equation}
We can use $R_{cl}$ to define the dimensionless variable 
\begin{equation}
	u = r / R_{cl}.
\end{equation}
The energies are naturally measured in units of the chemical potential $\mu$, so we introduce 
\begin{equation}
	V = \mu \bar V.
\end{equation}
In terms of the new variables, the action at the saddle becomes
\begin{equation}
  \Lag = \mu^{5/2} \bqty*{ c_0 \pqty*{ 1 - \bar V}^{5/2} + c_1 \frac{1}{R_{cl}^2 \mu} (\nabla_u \bar V)^2 },
\end{equation}
which shows that in the large-charge expansion, which is equivalent to \(\mu \gg (\nabla V)^2/V^2\), the subleading term is controlled by the parameter
\begin{equation}
  \epsilon^2 = \frac{2}{R_{cl}^2 \mu} \, ,
\end{equation}
as already observed in~\cite{Hellerman:2023myh} (see also~\cite{Hellerman:2020eff} for a careful analysis of the scales involved in the problem).

In this parametrization the coefficient \(c_2\) does not contribute to the action at the saddle and does not enter \emph{e.g.} in the expression for the conformal dimension of the lowest operator of fixed charge.
Conversely, in the following we will show that it does contribute at the same order as \(c_1\) to the dynamics of the fluctuations.

It is convenient to rewrite \(\epsilon\) as the ratio of a low-energy scale and a high-energy scale, introducing \(\varpi\) such that
\begin{align}
  \varpi &= \frac{\sqrt{2\mu}}{R_{cl}}  &\text{and} &&  \epsilon &= \frac{\varpi}{\mu } \, .
\end{align}
This parameter generically depends both on the confining potential and the chemical potential.
For a power-law potential of the form \(V \propto r^{2k}\), we can write
\begin{equation}
  V = \mu \pqty*{\frac{r}{R_{cl}}}^{2k} = \frac{\varpi^{2k}}{2^k \mu^{k-1}} r^{2k} \, .
\end{equation}
In the special case of a harmonic trap \(V = \omega^2 r^2/2\), we have \(\varpi = \omega\) and the dependence on \(\mu\) of the low-energy scale drops out entirely.
In what follows, potentials with $k>1$ are referred to as ``superharmonic''.

\paragraph{The small parameter \(\eta\).}

Expanding the action up to second order in the field we obtain the action that encodes the small fluctuations which are the subject of this work:
\begin{multline}
  \label{eq:quadratic-action}
	\Lag^{(2)}[\pi] = -\frac{5}{8}c_0 \sqrt{\mu-V} \pqty*{2(\mu-V)(\nabla \pi)^2-3\dot\pi^2} \\
	+ c_1 \bqty*{\frac{(\nabla \dot \pi)^2 }{\sqrt{\mu-V}}  + \frac{\nabla V \cdot \nabla \dot \pi^2}{2(\mu-V)^{3/2}} + \frac{(\nabla V)^2\pqty*{2(\mu-V)(\nabla \pi)^2+3\dot\pi^2}}{8(\mu-V)^{5/2}}} \\
   +  c_2\sqrt{\mu-V} \pqty*{(\Delta \pi)^2 -3 (\nabla \otimes \nabla \pi)^2} \ .
\end{multline}
One qualitative difference with respect to the analysis of the ground state is that, together with the low and high energy scales \(\varpi\) and \(\mu\), a third scale appears, namely the typical energy of the fluctuations $q_0$, which we use to rescale the time variable: 
\begin{equation}
	\bar t = q_0 t.
\end{equation}
Expressed in terms of dimensionless quantities, the action now takes the form 
 \begin{multline}
	\Lag^{(2)}[\pi] = -\frac{5}{8}c_0 \frac{\mu^{3/2}}{R_{cl}^2} \sqrt{1 - \bar V} \pqty*{2  (1 - \bar V)(\nabla_u \pi)^2 - 3 \frac{q_{0}^2 R_{cl}^2}{\mu} \pqty*{\pdv{\pi}{\bar t}}^2} \\
%	+ c_1 \frac{\mu^{1/2}}{R_{cl}^4}\bqty*{ \frac{q_{0}^2 R_{cl}^2}{\mu} \frac{(\nabla_u \dot \pi)^2 }{\sqrt{1 - \bar V}} + \frac{q_{0}^2 R_{cl}^2}{\mu} \frac{\nabla_u \bar V \cdot \nabla_u \dot \pi^2}{2( 1 - \bar V)^{3/2}} +  \frac{(\nabla_u \bar V )^2}{8(1 - \bar V)^{5/2}} \pqty*{2  (1 - \bar V)(\nabla_u \pi)^2 + 3 \frac{q_{0}^2 R_{cl}^2}{\mu} \dot \pi^2}} \\
   +  c_2 \frac{\mu^{1/2}}{R_{cl}^4} \sqrt{1 - \bar V} \bqty*{(\Delta_u \pi)^2 -3 \pqty*{\nabla_u \otimes \nabla_u \pi}^2}  + c_1 \frac{\mu^{1/2}}{R_{cl}^4} \Bigg[
     \frac{(\nabla_u \bar V )^2}{4(1 - \bar V)^{3/2}} (\nabla_u \pi)^2  \\
    +  \frac{q_{0}^2 R_{cl}^2}{\mu} \pqty*{ \frac{(\nabla_u \pdv{\pi}/{\bar t})^2 }{\sqrt{1 - \bar V}} + \frac{\nabla_u \bar V \cdot \nabla_u \pqty*{\pdv{\pi}/{\bar t}}^2}{2( 1 - \bar V)^{3/2}} +  \frac{3 \pqty*{\pdv{\pi}/{\bar t} \nabla_u \bar V }^2 }{8(1 - \bar V)^{5/2}} } \Bigg] 
\ .
\end{multline}
Comparing the different terms we see that
\begin{itemize}
\item the terms proportional to $c_1$ and $c_2$ are both by $\mu R_{cl}^2$ parametrically smaller than the corresponding terms proportional to $c_0$, as it was the case of the action at the saddle;
\item The two terms proportional to $c_0$ scale in the same way if we take $q_0=\order*{\sqrt{\mu}/R_{cl}}$.
  The same is true for all the terms proportional to $c_1$ and $c_2$.
\end{itemize}
The conclusion of this analysis is that we have a consistent expansion if at leading order $q_0 \sim \sqrt{\mu}/R_{cl}$ and the controlling parameter is
\begin{equation}
	\eta = \frac{q_0}{\mu}.
\end{equation}
Then the leading term is controlled by the low-energy constant $c_0$ and the \ac{nlo} correction is proportional to $c_1$ and $c_2$, which contribute both at the same order. 

\paragraph{The small parameter \(\delta\).}
The above is however not the only expansion that we will need to use: even if we limit ourselves to the leading order in the \ac{eft}, the \ac{eom} can in general not be solved exactly.
They take the form 
\begin{equation}\label{eq:WKB-EOM}\frac{1}{3}\frac{2 \mu}{q_0^2R_{cl}^2} \nabla_u \pqty*{(1-\bar V)^{3/2}\nabla_u\pi} =\sqrt{1-\bar V}\pdv[order=2]{\pi}{\bar t}.
\end{equation}
Since the system is invariant under time translation we can always write the solution in the form
\begin{equation}
	\pi(t, \mathbf{u}) = e^{iq_0t} \pi(\mathbf{u}).
\end{equation}
The \ac{eom} then become 
\begin{equation} 
	\frac{2 \mu}{(q_0 R_{cl})^{2}} \nabla_u\pqty*{(1-\bar V)^{3/2}\nabla_u \pi(\mathbf{u})} + 3 \pi(\mathbf{u}) \sqrt{1-\bar V} = 0 .
\end{equation}
We now have a differential equation in which the leading derivative is multiplied by a coefficient \(\delta^2\), where we have defined
\begin{equation}
  \label{eq:delta-parameter}
	\delta = \frac{\sqrt{2\mu}}{q_0 R_{cl}} = \frac{\varpi}{q_0}\ .
\end{equation}
In the limit in which the typical energy of the fluctuations is much larger than the low-energy scale, $\delta\ll 1$ and the \ac{eom} lends itself to a \ac{wkb} expansion.
We now have our second control parameter.

\paragraph{Regimes.}

The final result of the above considerations is that we have identified three scales: \(\varpi\),  \(q_0\),  and \(\mu\).
We will assume scale separation:
\begin{equation}
	\varpi \ll q_0 \ll \mu.
\end{equation}

From these scales we can construct two dimensionless parameters to realize a double expansion:
\begin{align}
	\eta &= \frac{q_0}{\mu} & \text{controlling the \ac{eft}},\\
	\delta &= \frac{\varpi}{q_0} & \text{controlling the \ac{wkb} expansion.}
\end{align}
The product $\eta \delta =\varpi/\mu = \epsilon \ll 1$ is the parameter that controls the perturbative large-charge expansion of the ground state.

\begin{figure}
  \centering
  \begin{tikzpicture}
    % Define the gradient
    \shade[left color=red, right color=blue] (0,0) rectangle (10,.35);

    % Top labels in speech balloon boxes
    \node at (1.5, 1) (low) [text width=2cm, align=left] {low energy (\acs{lo} \acs{eft})};
    \node at (8.5, 1) (high) [text width=3.6cm, align=right] {\mbox{high energy} (\acs{lo} \acs{wkb})};
    \node at (5, 1) (linear) []  {linear zone};

    % Bottom labels
    \node at (-2,-0.5) [text width=2cm, align=right] {fluctuation energy \(q_0\)};
    \draw[-latex] (-1,-.5) -- (11,-.5);
    \node at (0,-0.5) [fill=white] {$\varpi$};
    \node at (5,-0.5) [fill=white] {$\sqrt{\varpi \mu}$};
    \node at (10,-0.5) [fill=white] {$\mu$};

 %   \node at (1,-1.25) {\usebox{\lowenergyplot}};
 %   \node at (5,-1.25) {\usebox{\linearplot}};
 %   \node at (9,-1.25) {\usebox{\highenergyplot}};
\end{tikzpicture}
  \caption[Scale separation]{Scale separation for the fluctuation energy \(q_0\). The dynamics is controlled by the low-energy scale \(\varpi\), fixed by the confining potential and the chemical potential \(\mu\), acting as a high-energy scale. In the intermediate region (\(q_0 = \order{\sqrt{\varpi \mu}}\)), the physics is well described by the physical optics approximation for the \acs{lo} \acs{eft} and the measured dispersion is linear. At lower energies \(\varpi \ll q_0 \ll \sqrt{\varpi \mu} \), higher orders in the \acs{wkb} expansion are needed. At higher energies \( \sqrt{\varpi \mu} \ll q_0 \ll \mu \), higher orders in the \acs{eft} are needed.}
  \label{fig:energy-scales}
\end{figure}
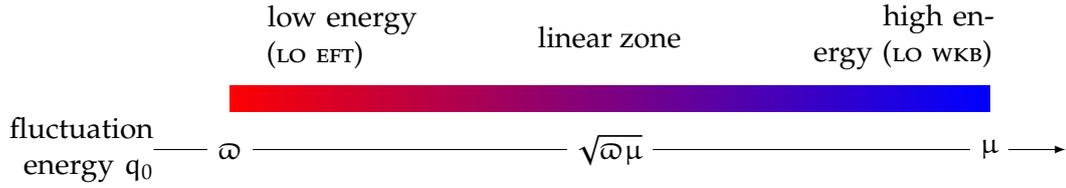

We can distinguish two regimes: 
\begin{itemize}
\item If $\eta \ll \delta \ll 1$, \emph{i.e.} $\varpi \ll q_0 \ll \sqrt{\varpi \mu}$ we are in the low-energy regime in which the dynamics is described by the \ac{lo} \ac{eft} and higher terms in the \ac{wkb} expansion.
\item If $\delta \ll \eta \ll 1$, \emph{i.e.} $\sqrt{\varpi \mu} \ll q_0 \ll \mu$ we are in the high-energy regime in which we can limit ourselves to the leading \ac{wkb} terms but need to add higher orders in the \ac{eft}.
\end{itemize}
The two regimes are separated by a linear zone where $q_0 \approx \sqrt{\varpi \mu}$ in which one can use the \ac{lo} \ac{eft} and \ac{lo} \ac{wkb} (see Fig.~\ref{fig:energy-scales}).

\paragraph{The \ac{wkb} expansion.}
{}\reversemarginpar\marginnote{\usebox{\lowenergyarrow}}%

As we have seen above, in general we can only find an approximate solution to the \ac{eom} using the \ac{wkb} expansion.
Here, we provide more detail.
For ease of exposition, we consider a spherically symmetric potential $V=V(r)$ and focus on the low-energy regime, where we can confine ourselves to the \ac{lo} in the \ac{eft}.
Taking the time-translation and $SO(3)$ rotational symmetries into account, the field takes the form
\begin{equation}
\label{equ:wave_decomposition}
  \pi(t,\mathbf{r}) = e^{i q_0 t} \pi(r) Y_{\ell m}(\vartheta, \phi).
\end{equation}
The \ac{eom} \eqref{eq:WKB-EOM} now separates and for the radial component in terms of the dimensionless variable $u$, we find 
\begin{equation}
  \delta^2 \odv{}{u}\pqty*{u^2 \pqty*{1 - \bar V}^{3/2} \odv{}{u}\pi(u)} - \delta^2 \ell \pqty{ \ell + 1} \pqty{ 1 - \bar V}^{3/2} \pi(u)+ 3 u^2 \sqrt{1 - \bar V} \pi(u) = 0.
\end{equation}
As we have observed above, the leading derivative is controlled by the parameter \(\delta\).
In the limit $\delta \to 0$, we can look for a \ac{wkb} perturbative solution of the form
\begin{equation}\label{eq:WKB}
  \pi(u) = \exp*[\frac{i}{\delta} \scS_0(u) + \scS_1(u) + i \delta \scS_2(u) + \delta^2\scS_3(u) + \dots ].
\end{equation}
The factors of \(\sqrt{-1}\) in the ansatz are chosen such that all the equations are real.
It is then apparent that the even functions \(\scS_{2n}\) control the frequency of the oscillations, while the odd ones \(\scS_{2n+1} \) control the amplitude of the fluctuations \(\pi(u)\).

Expanding order by order, we find a hierarchy of algebraic equations for the derivatives \(\scS_p'(u)\).
The leading (eikonal) equation is
\begin{equation}
   \pqty*{ 1 - \bar V(u) } \scS_0'(u)^2 = 3 \, ,
\end{equation}
and can be solved immediately for \(\scS_0' \).
The next orders are
\begin{gather}
  \pqty*{3 u \bar V' - 4 \pqty*{ 1- \bar V}} \scS_0' - 2 u \pqty*{1 - \bar V} \pqty*{ \scS_0'' + 2 \scS_0' \scS_1' } = 0 ,\\ 
  \pqty*{3 u \bar V' - 4 \pqty*{ 1- \bar V}} \scS_1' - 2 u \pqty*{1 - \bar V} \pqty*{ \scS_1'' - 2 \scS_0' \scS_2'  + (\scS_1')^2 } + \frac{2}{u} \ell \pqty{ \ell + 1 } (1 - \bar V) = 0 .
\end{gather}
The angular momentum \(\ell\) only appears in the equation for \(\scS_2\), and the general equation for \(\scS_{p+1}'\), \(p = 2, 3, \dots \) is
\begin{equation}
    \pqty*{3 u \bar V' - 4 \pqty*{ 1- \bar V}} \scS_p' - 2 u \pqty*{1 - \bar V} \pqty*{ \scS_p'' + \sum_{k=0}^{\floor{p/2}} I_{p,k} \scS_k' \scS_{p+1-k}' - \sin*(\tfrac{ 3 p \pi}{2}) (\scS'_{(p+1)/2})^2} = 0 \, ,
\end{equation}
where \(I_{p,k} = 1 - (-1)^k + (-1)^p + (-1)^{k+p}\).

The \ac{wkb} expansion can also be understood as a gradient expansion for the potential.
From the structure of the equations we see that \(\scS_0\) and \(\scS_1\) only depend on \(V\), while \(\scS_2\) and \(\scS_3\) depend on the first and second derivatives \(V' \) and \(V''\), and so on.

The same analysis can be performed for the \ac{wkb} expansion of the \ac{nlo} \ac{eom}, which must be used in the high-energy regime.
Thanks to scale separation, the result is that at each order in the \ac{wkb} hierarchy there will be a new term, suppressed by \(\eta = q_0/\mu\).
We will discuss this in detail in Section~\ref{sec:high-energy-regime}.

\section{Fluctuation spectrum}
\label{sec:fluctuation-spectrum}

Our aim in this section is to compute the spectrum of the system in
presence of a trapping potential \(V(r)\), \emph{i.e.} the allowed
values of the energy of small fluctuations.  The natural expectation
is for the spectrum to be discrete, because of the confining nature of
the potential.  Here we make this mathematically precise, showing that
the operator that controls the fluctuation is self-adjoint both at
\ac{lo} and \ac{nlo} in the \ac{eft} expansion.  The \ac{eom} then
turns into a Sturm--Liouville problem (\ac{lo}) and its fourth-order
generalization (\ac{nlo}).

%%%%%%%%%%%%%%%%%%%%%%%%%%%%%%%%%%%%%%%%%%%%%%%%%%%%%
\subsection{The linear regime}
\label{sec:linear-regime}
{}\reversemarginpar\marginnote{\usebox{\midenergyarrow}}[-1cm]%
% \marginfigure{  \begin{tikzpicture}[scale=.2]
%     % Define the gradient
%   \shade[left color=red, right color=blue] (0,0) rectangle (10,1);
%   \draw[-{Triangle[width=18pt, length=10pt]}, line width=10pt, blue!50!red] (5,4) -- (5,1) ;
% \end{tikzpicture}}[-1cm]

Let us start by considering the regime
\begin{equation}
  \varpi \ll q_0 \ll \mu \, ,
\end{equation}
which---for reasons that will become clear in the following---we call the \emph{linear regime}.
In this approximation, both control parameters \(\delta = \varpi/q_0\) and \(\eta = q_0 / \mu\) are very small, and the dynamics is well described by the \ac{lo} \ac{eft}, whose \ac{eom} we can solve in the physical optics approximation (defined below).

\paragraph{Structure of the problem.}

The \ac{lo} linearized \ac{eom} reads 
\begin{equation}
	3(\mu-V)^{1/2}\ddot{\pi} - \nabla(2(\mu-V)^{3/2}\nabla \pi) =0.
\end{equation}
The general structure becomes clearer if we rewrite the equation for a more general action $\Lag$ which is only a function of the Galilean-invariant object 
\begin{equation}
	X = \del_t \theta -\tfrac{1}{2} (\nabla\theta)^2 -V.
\end{equation}
In our case, $\Lag[X] = c_0 X^{5/2}$. So we consider the equation
\begin{equation}
	\Lag'' [\mu- V] \ddot{\pi} - \nabla(\Lag'[\mu- V]\nabla \pi) = 0.
\end{equation}
If the potential has spherical symmetry $V=V(r)$, the system is invariant under time translation and $SO(3)$ rotations, so we can write $\pi(t,\mathbf{r})$ as (\ref{equ:wave_decomposition}). The \ac{eom} separates and reduces to an \ac{ode} for the radial mode $\pi(r)$:
\begin{equation}
	\frac{\dd{}}{\dd{r}}\pqty*{r^2 \Lag'[\mu- V] \frac{\dd{}}{\dd{r}}\pi } - \Lag'[\mu- V] \ell(\ell+1) \pi = - r^2 \Lag'' [\mu- V] q_0^2 \pi.
\end{equation}
This is a standard \emph{Sturm--Liouville problem} on the interval $0<r<R_{cl}$ with eigenvalue $q_0^2$ and weight function 
\begin{equation}
	w(r) = r^2 \Lag''[\mu-V(r)].
\end{equation} 
It follows that the spectrum is discrete. For each allowed value of $q_0^2$ and fixed \(\ell\) there is a single eigenfunction $\pi_{n,\ell}$, and these eigenfunctions form an orthonormal basis for the Hilbert space $L^2$ with inner product
\begin{equation}
	\braket{n|m} = \int_0^{R_{cl}} w(r) \dd{r} \pi_n(r) \pi_m(r) = \delta_{nm}.
\end{equation}
The weight function $ w(r)$ has two parts.
The $r^2$ factor is purely geometrical and corresponds to a three-dimensional system. The non-trivial part results from the fact that we are considering fluctuations in a non-homogeneous medium.

A physical interpretation of the weight can be obtained as follows.
For $V=0$, the compressibility sum rule reads
\begin{equation}
	\odv[order=2]{\Lag}{\mu} = \frac{\rho}{c_s^2},
\end{equation}
where $\rho$ is the density, $c_s$ is the speed of sound, and we have identified the action at the saddle with the pressure as a function of the chemical potential.
For the case of a confining potential, we can identify the weight function with the ratio of density and the square of the speed of sound which are now both position dependent:
\begin{equation}
  w(r) = r^2 \frac{\rho(r)}{c_s^2(r)}.
\end{equation}
This is precisely the same relation that one would have obtained in the \ac{lda} by trading the chemical potential \(\mu\) in the \(V = 0\) expressions for the combination \(\mu - V(r)\).
In our case, it is however the result of a calculation and not of having imposed an approximation.

\bigskip

The same analysis can be performed for a cylinder geometry, where the system is confined in $x^2+y^2 < R_{cl}^2$ and $0<z<L$, and the potential depends only on $r=\sqrt{x^2+y^2}$.
In this case, the fluctuations take the form
\begin{equation}
	\pi(t,\mathbf{r}) = e^{iq_0t}\sin*(2\pi\frac{n_z z}{L}) e^{in_\vartheta\vartheta}\pi(r),
\end{equation}
where $n_z$ and $n_\vartheta$ are integers.
The \ac{eom} is again a Sturm--Liouville problem for the radial field:
\begin{equation}
	\frac{\dd{}}{\dd{r}}\pqty*{ r \Lag'[\mu-V] \frac{\dd{}}{\dd{r}}\pi} - \Lag'[\mu-V]  \pqty*{\frac{n_{\vartheta}^2}{r}+\frac{r}{L^2}(2\pi n_z)^2} \pi = - r \Lag''[\mu-V] q_0^2 \pi.
\end{equation}
The corresponding weight function only differs in the geometric factor since now the problem is effectively two-dimensional:
\begin{equation}
	w(r) = r \Lag''[\mu-V] = r \frac{\rho(r)}{c^2(r)}.
\end{equation}
The geometry introduces yet another new scale \(L\), which generically changes the \ac{eft} description.

\paragraph{Physical optics approximation for the \ac{lo} \ac{eft}.}

Having understood the general structure of the problem, we can try to find an approximate solution.
In the linear regime, it is appropriate to use the \emph{physical optics approximation} (the two leading \ac{wkb} equations).
The equations to solve are
\begin{align}
 {}& \pqty*{ 1 - \bar V(u) } \scS_0'(u)^2 = 3 \, ,
  \label{eq:eikonal}\\
  {}&  \pqty*{3 u \bar V' - 4 \pqty*{ 1- \bar V}} \scS_0' - 2 u \pqty*{1 - \bar V} \pqty*{ \scS_0'' + 2 \scS_0' \scS_1' } = 0 \, ,
  \label{eq:transport}
\end{align}
which are, respectively, the \emph{eikonal} and \emph{transport} equations.

One finds immediately
\begin{align}
  \scS_0(u) &= \pm \sqrt{3}  \int_{u_0}^u \frac{\dd{w}}{\sqrt{ 1 - \bar V(w)}}\,, \\
  \scS_1(u) &= - \log*(u \sqrt{ 1 - \bar V(u)}) + k_1 \, , 
\end{align}
where \(u_0\) and \(k_1\) are constants.
The solution to the \ac{eom} at this order is then
\begin{equation}
  \pi(u) = \frac{ D_+ \exp*[i \sqrt{3}\frac{q_0}{\varpi} \int_{u_0}^u  \frac{\dd{w}}{\sqrt{ 1 - \bar V(w)}}] + D_- \exp*[-i \sqrt{3} \frac{q_0}{\varpi} \int_{u_0}^u  \frac{\dd{w}}{\sqrt{ 1 - \bar V(w)}}]}{u \sqrt{1 - \bar V(u)}} \, ,
\end{equation}
where \(D_{\pm}\) are constants.

For general values of the parameters there is a singularity in \(u = 0\) (the center of the drop) and at \(u = 1\), where \(\bar V(u) = 1\) (the edge of the drop).
These singularities are not physical since we want to describe small fluctuations.
Eliminating them will fix the form of the spectrum, that we expect to be discrete due to the presence of a confining potential.

Around the origin \(u = 0\), \(\scS_0\) is given by
\begin{equation}
  \scS_0(u) \underset{u \to 0}{\sim} \pm \sqrt{3} \left( \int_{u_0}^0 \frac{\dd{w}}{\sqrt{1 - \bar V(w)}} + \frac{ u}{\sqrt{1 - \bar V(0)}} + \dots \right),
\end{equation}
so that the field \(\pi\)  is
\begin{equation}
  \pi(u) \underset{u \to 0}{\sim} \frac{ D_+ \exp*[i \sqrt{3}\frac{q_0}{\varpi} \int_{u_0}^0  \frac{\dd{w}}{\sqrt{ 1 - \bar V(w)}}] + D_- \exp*[-i \sqrt{3} \frac{q_0}{\varpi} \int_{u_0}^0  \frac{\dd{w}}{\sqrt{ 1 - \bar V(w)}}] }{u \sqrt{ 1 - \bar V(0)}}  + \order{u^0},
\end{equation}
which is singular unless the quantity in the bracket vanishes.
Setting, without loss of generality, \(u_0 = 0\) and \(D_+ = - D_- = D/(2i)\) we obtain a solution that is regular in \(u = 0\):
\begin{equation}
  \label{eq:physical-optics-fluctuations}
  \pi(u) = D \frac{ \sin*( \sqrt{3} \frac{q_0}{\varpi} \int_0^u \frac{\dd{w}}{\sqrt{1 - \bar V(w)}} )}{u \sqrt{1 - \bar V(u)}} \, .
\end{equation}
We still need to deal with the singularity in \(u = 1\).
Asking for the \(\sin\) to vanish, we obtain a quantization condition for \(q_0\):
\begin{align}
  \label{eq:q0-quantization}
   \frac{1}{\delta} \scS_0(1) = \frac{q_0}{\varpi} \sqrt{3} \int_0^1 \frac{\dd{w}}{\sqrt{1 - \bar V(w)}} = n \pi, && n \in \setZ .
\end{align}
At leading order the states in the spectrum are equidistant.
This was to be expected, because at leading order in the \ac{wkb} expansion the system behaves qualitatively as if the potential were constant (see Eq.~\eqref{eq:spherical-box-spectrum}).

Our approximation is consistent for \(q_0 \gg \varpi\), which is the same as \(n \gg 1\).
From the general analysis in the previous section, we know that these solutions form an orthogonal basis, which can be made orthonormal with an appropriate choice of the constant \(D\):
\begin{align}
  \int \dd{u} u^2 \sqrt{1 - \bar V(u)} \pi_{r}^n(u) \pi_{r}^m (u) = \delta_{mn} 
\end{align}
if
\begin{equation}
  \label{eq:fluctuation-normalization}
  \frac{1}{D^2} = \frac{1}{2}\int_0^1 \frac{\dd{w}}{\sqrt{1 - \bar V(w)}} \, .
\end{equation}

For a power-law potential of the form
\begin{align}
  V(r) &= \mu \pqty*{\frac{r}{R_{cl}}}^{2k}, & R_{cl} &= \frac{\sqrt{2\mu}}{\varpi} \, ,
\end{align}
the reduced potential is just \(\bar V = u^{2k}\). In this case, we can solve the integral in Eq.~\eqref{eq:physical-optics-fluctuations} analytically and obtain the spectrum equation
\begin{equation}%
  \label{eq:quantizationLOLO}
  q_0(n) = \frac{2}{\sqrt{3} \binom{1/(2k)}{1/2}} \varpi n . % n  \omega \sqrt{\frac{\pi}{3}} \frac{\Gamma(\frac{1}{2} + \frac{1}{2k})}{\Gamma(1 + \frac{1}{2k})} n .
\end{equation}
%and the energy scale is \(\varpi = (\mu \omega)^{1/2} (\omega/\mu)^{1/(2k)}\).
The slope of $q_0$ becomes steeper for flatter potentials (larger values of \(k\)) and varies between \(q_0 (n)= 2 n \omega/\sqrt{3}\) for the harmonic potential (\(k = 1\)), and \(q_0(n) = \pi n \varpi / \sqrt{3}\) in the \(k \to \infty\) (spherical box) limit.\\
The fluctuation can be written explicitly in terms of hypergeometric functions:
\begin{equation}
    \label{eq:physical-optics-superharmonic-oscillator}
   \pi(u) = \sqrt{2k \binom{\sfrac{1}{2k} - \sfrac{1}{2}}{\sfrac{1}{2}}} \frac{ \sin*( \frac{2 n}{\binom{1/(2k)}{1/2}} u \  \pFq{2}{1}{\frac{1}{2},\frac{1}{2k}}{1+\frac{1}{2k}}{u^{2k}} )}{u \sqrt{1 - u^{2k}}} \, ,
\end{equation}
which for the harmonic oscillator is simply%
\footnote{The \ac{eom} for the fluctuations in the \ac{lo} \ac{eft} with a harmonic potential can be solved exactly and our \ac{lo} result is consistent with the appropriate limit of the exact solution (see Section~\ref{sec:fluctuations-harmonic-potential}).}
\begin{equation}
  \label{eq:physical-optics-harmonic-oscillator}
   \pi(u) = \frac{2}{\sqrt{\pi}} \frac{ \sin*( 2n \arcsin(u) )}{u \sqrt{1 - u^2}} \, .
\end{equation}
 
\begin{figure}
  \centering
  \includegraphics[width=.75\textwidth]{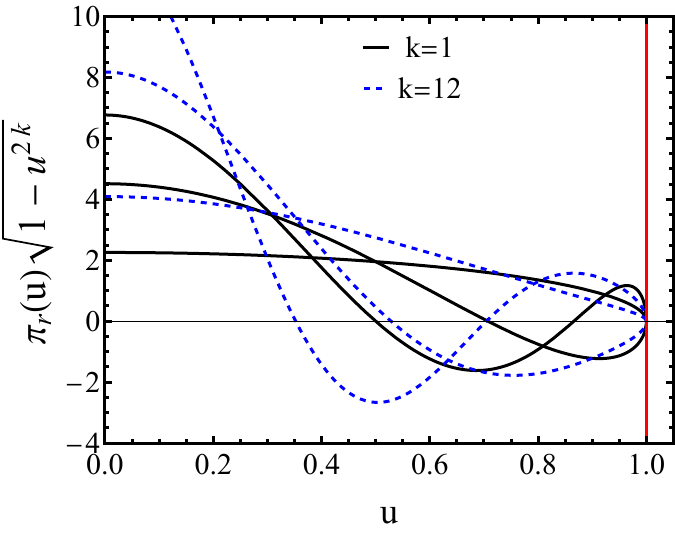}
  \caption{Fluctuations of the charge density, taken from Eq.~(\ref{eq:physical-optics-superharmonic-oscillator}), with $k=1$ (solid lines) and $k=12$ (dashed lines)
    for the first three levels. The level counts the number of nodes.}
  \label{fig:wf-linear-regime}
\end{figure}

The fluctuations of the charge density are illustrated in Fig.~\ref{fig:wf-linear-regime}.
Note that the result does not depend on the angular momentum \(\ell\) of the fluctuations.
To see the effect of the angular momentum we need to add the next order in the \ac{wkb} expansion, see Section~\ref{sec:low-energy-regime}.

%%%%%%%%%%%%%%%%%%%%%%%%%%%%%%%%%%%%%%%%%%%%%%%%%%%%%
\subsection{High energy regime}
\label{sec:high-energy-regime}
{} \reversemarginpar\marginnote{\usebox{\highenergyarrow}}[-1cm]

Let us now consider the high-energy region  
\begin{equation}
	\frac{\varpi}{q_0} \ll \frac{q_0}{\mu} \ll 1
\end{equation}
in which we have to take into account the \ac{nlo} corrections to the \ac{eft}---which are controlled by \(q_0 / \mu\) ---but we can limit ourselves to the physical optics terms in the \ac{wkb} expansion.

The \ac{eom} for the fluctuations described by the action in Eq.~\eqref{eq:quadratic-action} are of fourth order.
Assuming a radially-symmetric potential, the equations separate and the \ac{eom} for the radial function \(\pi(u)\) takes the form
\begin{equation}
 \delta^4  \odv[order=2]{}{u} \pqty*{ p(u)  \odv[order=2]{\pi(u)}{u}} -  \delta^2  \odv{}{u} \pqty*{s(u) \odv{\pi(u)}{u}} =  w(u) \pi(u),  
\end{equation}
where
\begin{align}
  p(u)&= -4 u^2  \sqrt{1-\bar V}q_0^2 c_2 ,\\
 s(u)&= - 5u^2\mu^2(1-\bar V)^{3/2}c_0 +u^2\left(\frac{\delta^2 q_0^2 (\bar V')^2}{2(1-\bar V)^{3/2}}+ \frac{4 q_0^2}{\sqrt{1-\bar V}}\right)c_1 -4q_0^2\delta^2 c_2\left(4\sqrt{1-\bar V} -\frac{u \bar V'}{\sqrt{1-\bar V}}\right) ,\\
  w(u)&=  -15 u^2 \mu^2 \pqty*{1 - \bar V}^{1/2} c_0  - 8  q_0^2\frac{\odv{}{u} \pqty*{ u^2  \odv{}{u}\pqty*{1- \bar V}^{1/4}}}{\pqty*{1 - \bar V}^{3/4}} \delta^2 c_1.
\end{align} 
We have omitted the $\ell$-dependence, as it contributes only at higher order in the \ac{wkb} approximation. Imposing appropriate boundary conditions, the fourth-order differential operator is selfadjoint, so we have a generalization of the Sturm--Liouville problem encountered at \ac{lo} in Sec.~\ref{sec:linear-regime}.
This means that the spectrum of \(q_0(n)^2\) remains discrete, and there is a basis of eigenvectors \(\pi_n(r)\) for the Hilbert space of fluctuations with weight \(w(r)\).

The highest derivative term is again controlled by the small parameter \(\delta = \varpi/q_0\), and we can look for a perturbative \ac{wkb} solution.
As pointed out above, in this regime it is sufficient to correct the eikonal and transport equations in Eq.~\eqref{eq:eikonal} and Eq.~\eqref{eq:transport} which become
\begin{align}
  &{} 5 \mu^2 (1-\bar V) c_0\left( 3-(1-\bar V) {\scS_0'}^2 \right)+ 4 c_1 q_0^2 {\scS_0'}^2 -4 c_2 q_0^2  (1-\bar V){\scS_0'}^4=0  ,\\ 
  &{}\begin{multlined}[.9\textwidth]
   5 c_0 \mu^2 \pqty*{1 - \bar V} \bqty*{3u  \bar V'\scS_0'  -\pqty*{1 - \bar V} \pqty*{4 \scS_0' \pqty*{1+u \scS_1'} + 2 u  \scS_0''}}\\
   +4 c_1 q_0^2 \pqty*{ \scS_0' \pqty*{4 + 4 u \scS_1' + \frac{u \bar V}{1 - \bar V}}+ 2 u \scS_0''} \\
    + 8 c_2 q_0^2 {\scS_0'}^2 \bqty*{u   \bar V' \scS_0'- \pqty*{1 - \bar V} \pqty*{4\scS_0' \pqty*{1+u \scS_1'} + 6 u \scS_0''}} = 0 .
  \end{multlined}
\end{align}
Expanding at the appropriate order in \(q_0 / \mu\), we find the solution:
\begin{align}
  \label{eq:NLO-EFT-LO-WKB}
	\scS_0 &= \pm \sqrt{3} \int_{u_0}^u \dd{w} \frac{1}{\sqrt{1-\bar V(w)}}\pqty*{1+\frac{2}{5c_0}\frac{c_1-3c_2}{\pqty*{1-\bar V(w)}^2}\frac{q_0^2}{\mu^2}+\order*{\frac{q_0^4}{\mu^4}} }, \\
	\scS_1 &= k_1 - \log*(u\sqrt{1-\bar V}) +\frac{c_1 - 9
    c_2}{5 c_0(1-\bar V)^2}\frac{q_0^2}{\mu^2}+\order*{\frac{q_0^4}{\mu^4}} .
\end{align}
%Studying the system at higher order in \ac{wkb}, we have observed that (for $\ell \neq 0$) the singularity at the origin was effectively moved by an amount of order $\mathcal{O}(\omega^2/q_0^2)$.\todo{rewrite. The order of subsections has changed.}
%We have a similar interpretation here for the new term depending on $c_1$ and $c_2$ in $\scS_1$:
%the singularity corresponding to $\bar V = 1$ is shifted inwards by an amount proportional to $\order{\omega^2/q_0^2}$.
At leading order, we have seen that the solution is in general singular at the edge of the droplet, $u=1$, and picking only regular solutions leads to a discrete spectrum.
Adding the next term in the \ac{wkb} expansion can shift the singularity inwards by an amount proportional to $\order{q_0^2/\mu^2}$.
This is reminiscent of the fact that the size of the droplet decreases, compared to the \ac{lo} one, when including \ac{nlo} corrections to the \ac{eft}.
This effect becomes manifest if we write
\begin{equation}
	\scS_1(u) = k_1 - \log u - \frac{1}{4}\log\pqty*{(1-\bar V(u))^2-\frac{4}{5c_0}(c_1-9c_2)\frac{q_0^2}{\mu^2}}+\order*{\frac{q_0^3}{\mu^3}}.
\end{equation}
There are however two possible cases, depending on the sign of the combination $c_1-9c_2$. 

\paragraph{Positive sign.} If the sign is positive, which is what is found in mean-field theory (large-$N$ ~\cite{Ma_es_2009,Kurkjian_2016,Hellerman:2023myh} ), the  $\bar V=1$ singularity is shifted to $u = u_1 $, where $u_1$ is the solution to
\begin{equation}
	\pqty*{1-\bar V(u_1)}^2 - \frac{4}{5c_0}(c_1 - 9c_2)\frac{q_0^2}{\mu^2} =0.
\end{equation}
Once more, we are looking for regular solutions for our small oscillations.
Imposing regularity at the origin, we find:
\begin{equation}
  \pi(u) = D \frac{\sin*(\sqrt{3}\frac{q_0}{\varpi} \int_0^u \dd{w} \frac{1}{\sqrt{1-\bar V(w)}}\pqty*{1+\frac{2}{5c_0}\frac{c_1-3c_2}{\pqty*{1-\bar V(w)}^2}\frac{q_0^2}{\mu^2}  })
}{u \pqty*{(1-\bar V(u))^2 -\frac{4}{5c_0}(c_1-9 c_2)\frac{q_0^2}{\mu^2}}^{1/4}},
\end{equation}
and regularity at $u = u_1$ leads to the quantization condition:
\begin{equation}
	\sqrt{3}\frac{q_0}{\varpi} \int_0^{u_1} \dd{w} \frac{1}{\sqrt{1-\bar V(w)}}\pqty*{1+\frac{2}{5 c_0}\frac{c_1-3c_2}{(1-\bar V(w))^2}\frac{q_0^2}{\mu^2}} = n\pi.
\end{equation}
In the case of a power-law potential $\bar V(u) = u^{2k}$, we can solve the integral explicitly in terms of gamma functions.
First, we need to solve for the location of the singularity:
\begin{equation}
\label{eq:NLO-edge-positive-sign}
	u_1= 1- \frac{1}{2k} \sqrt{\frac{4}{5c_0}(c_1-9c_2)}\frac{q_0}{\mu}.
\end{equation}
The integral in the spectrum equation can be solved explicitly to find the first correction to \ac{lo} in Eq.~\eqref{eq:quantizationLOLO}:
\begin{equation}
	\frac{q_0}{\varpi}\pqty*{\sqrt{3}\frac{\pi}{2} \binom{1/(2k)}{1/2}+ \frac{51c_2-5c_1}{\sqrt{6}(5c_0)^{1/4} k (c_1-9c_2)^{3/4}} \sqrt{\frac{q_0}{\mu}}} = n\pi \ ,
\end{equation}
which can be inverted to give
\begin{equation}
  \frac{q_0(n)}{\varpi} = \frac{2}{\sqrt{3} \binom{1/(2k)}{1/2}} n + \frac{4 \pqty*{5 c_1 - 51 c_2}}{3^{7/4} \pi k (5 c_0)^{1/4} \pqty*{c_1 - 9 c_2}^{3/4} {\binom{1/(2k)}{1/2}}^{5/2}} \sqrt{\frac{\varpi}{\mu}} n^{3/2} + \order*{\frac{\varpi}{\mu}n^2} .
\end{equation}

\paragraph{Negative sign.} If the sign of $c_1-9c_2$ is negative, as is found in the $\epsilon$ expansion~\cite{Rupak_2009}, the singularity remains in $u=1$. 
However, the \ac{eft} expansion breaks down close to $u=1$ because the \ac{nlo} term becomes larger than the \ac{lo} one. 
We can only trust our approximation up to distances of the order of $u \lessapprox 1-\order{q_0/\mu}$, which is consistent with the value of $u_1$ we found in the other case in Eq.~\eqref{eq:NLO-edge-positive-sign}.
The final result is that the spectrum takes again the form
\begin{equation}
  \frac{q_0(n)}{\varpi} = \frac{2}{\sqrt{3} \binom{1/(2k)}{1/2}} n + \lambda \sqrt{\frac{\varpi}{\mu}} n^{3/2} + \order*{\frac{\varpi}{\mu}n^2} ,
\end{equation}
where now $\lambda$ is some new parameter depending on the edge physics, which is not accessible within our approximation.

The \ac{nlo} in the \ac{eft} is sensitive to the effects from the cloud edge. This is expected, since the control parameter of the \ac{eft} is ultimately the charge density, which vanishes at the edge.
Already in the study of the ground-state energy, this leads to the appearance of new fractional powers in the charge~\cite{Kravec:2018qnu,Hellerman:2021qzz}.
Here, we see that the leading order of the spectrum equation receives a correction that scales like $n^{3/2}$, with a coefficient that is computable in terms of the low-energy constants in the case of $c_1-9c_2>0$ or requires the inclusion of edge terms for $c_1-9c_2<0$.

%%%%%%%%%%%%%%%%%%%%%%%%%%%%%%%%%%%%%
\subsection{No potential}
\label{sec:no-potential}
{}\reversemarginpar\marginnote{\usebox{\highenergyarrow}}[-1cm]%

As a limit case of the high-energy regime, we can consider the case of zero potential \(V = 0\), which amounts to \(\varpi = 0\). Here, the only remaining expansion parameter is \(\eta = q_0/\mu\).
We can also think of this as the limit $k \to \infty$ for the confining potential $u^{2k}$. In order to retain a confining effect, we need to impose Dirichlet boundary conditions at $r= R_{cl}$, which results in a confining bucket.

The quadratic action for the Goldstone $\pi(t, \mathbf{r})$ in Eq.~\eqref{eq:quadratic-action} reduces to
\begin{equation}
  \Lag^{(2)}[\pi] = -\frac{5}{8}c_0 \sqrt{\mu} \pqty*{2\mu(\nabla \pi)^2 - 3 \dot \pi^2} +\frac{c_1}{\sqrt{\mu}} \pqty*{\nabla \dot \pi}^2 + c_2\sqrt{\mu} \pqty*{(\Delta \pi)^2 -3 (\nabla \otimes \nabla \pi)^2} \, ,
\end{equation}
and the corresponding \ac{eom} is
\begin{equation}
  \label{eq:eomnopot}
  5 c_0 \mu \pqty*{3 \ddot \pi - 2\mu \Delta \pi} - 8c_1 \Delta \ddot \pi + 16 c_2\mu \Delta^2\pi = 0,
\end{equation}
where $\Delta^2$ is the biharmonic operator 
\begin{equation}
  \Delta^2\pi = \sum_{i,j=1}^3 \del_i\del_i\del_j\del_j \pi.
\end{equation}

In this configuration the symmetries of the problem change and we have the full Poincaré group \(ISO(1,3)\); moreover the spectrum is continuous, so the natural Ansatz is a simple plane wave,
\begin{equation}
  \pi(t, \mathbf{r}) = \exp*[ i q_0 t + i \mathbf{q} \cdot \mathbf{r}] \, .
\end{equation}
Substituting this ansatz into Eq.~\eqref{eq:eomnopot}, we find the dispersion relation for the plane wave:
\begin{equation}
  5 c_0 \mu (-3 q_0^2 + 2 q^2\mu) - 8c_1 q_0^2q^2 + 16 \mu c_2 q^4 = 0.
\end{equation}
Equivalently, solving for $q_0$, we find
\begin{equation}%
\label{eq:no-trap-dispersion}
  q_0 (q)= \sqrt{\frac{2\mu}{3}}q- \frac{4}{15c_0}\sqrt{\frac{2}{3\mu}}(c_1 - 3 c_2) q^3 + \mathcal{O}(\mu^{-3/2}),
\end{equation}
which can be parametrized as 
\begin{equation}
  q_0(q) = c_s q \pqty*{1 + \frac{\gamma}{8 c_s^2}q^2}
\end{equation}
with
\begin{align}%
  \label{eq:EFT-dispersion-parameters}
  c_s &= \sqrt{\frac{2\mu}{3}}, & \gamma & = \frac{64}{45} \frac{3c_2 - c_1}{c_0},
\end{align}
as already found above.
In this simple case, the Goldstone nature of the fluctuation becomes completely manifest. Since the low-energy scale is zero, we can see here that $q_0$ is zero when $q=0$ (which is not the case when $q=0$ is not accessible with the \ac{eft} due to the presence of a potential). The value of $c_s$ is completely fixed by scale invariance. For a flat potential at criticality, the low-energy spectrum must be linear with the slope fixed by the value of $\mu$. We will find this slope again when computing the dynamical structure factor in Section~\ref{sec:dynamic-structure-factor}.

\bigskip
To make contact with the \ac{wkb} solution, we could have also looked for a spherically-symmetric solution of the form 
\begin{equation}%
    \pi_{q_0, q}(t,r) = e^{iq_0 t} \pi(r) 
\end{equation}
for which we would have found the \ac{eom} 
\begin{multline}
  c_0\pqty*{-\frac{5}{2}\mu^{3/2}\pi'' - 5\sqrt{\mu} \pqty*{\frac{\mu}{r}\pi' +\frac{3}{4}q_0^2\pi}} + 2\frac{c_1q_0}{\sqrt{\mu}} \pqty*{\pi'' + \frac{2}{r}\pi'} \\
  + 4 c_2\sqrt{\mu}\pqty*{\pi '''' + \frac{4}{r} \pi '''} = 0.
\end{multline}
This admits a regular spherical wave solution
\begin{equation}%
  \label{eq:spherical-wave}
    \pi_{q_0, q}(t, r) = e^{iq_0 t} \frac{\sin(q r)}{r \sqrt{\mu}},
\end{equation}
where \(q_0\) and \(q\) are related by the same dispersion relation as for the plane wave.

In the case of a spherical box in which we impose Dirichlet boundary conditions \(\pi(R_{cl})  =0\) at \(R_{cl} = \sqrt{2\mu}/\varpi\), the spectrum is discrete and reads:
\begin{equation}
  \label{eq:spherical-box-spectrum}
  \frac{q_0}{\varpi} = \frac{\pi}{\sqrt{3} } n  \, ,
\end{equation}
which is the large-\(k\) limit of the relation that we have found in Eq.~\eqref{eq:quantizationLOLO} (see also Fig.~\ref{fig:low-energy-spectrum}).
This is consistent with our results from the \ac{wkb} expansion: the solution given in Eq.~\eqref{eq:physical-optics-fluctuations} reduces to this form for $\bar V=0$.
In this case, the eikonal approximation is exact, as the higher-order terms in the \ac{wkb} expansion take into account the effects of a potential.

\subsection{Low energy regime}
\label{sec:low-energy-regime}
{}\reversemarginpar\marginnote{\usebox{\lowenergyarrow}}[-1cm]%

To go beyond the linear approximation in the low-energy regime, we need to add higher-order corrections to the \ac{wkb} expansion.
We study the low-energy region, where
\begin{equation}
  \frac{q_0}{\mu} \ll \frac{\varpi}{q_0} \ll 1 \, ,
\end{equation}
where we can neglect the \ac{nlo} terms in the \ac{eft}.

The new equations to solve are
\begin{align}
   {}& \pqty*{3 u \bar V' - 4 \pqty*{ 1- \bar V}} \scS_1' - 2 u \pqty*{1 - \bar V} \pqty*{ \scS_1'' - 2 \scS_0' \scS_2'  + (\scS_1')^2 } + \frac{2}{u} \ell \pqty{ \ell + 1 } (1 - \bar V) = 0\, ,\\
   {}& \pqty*{3 u \bar V' - 4 \pqty*{ 1- \bar V}} \scS_2' - 2 u \pqty*{1 - \bar V} \pqty*{ \scS_2'' + 2 \scS_0' \scS_3' + 2 \scS_1' \scS_2'  } = 0\, .
\end{align}
The solution can be written in terms of a function of \(\bar V\):
\begin{align}
  \scS_2(u) &=  \frac{1}{4\sqrt{3}} \int_{u_0}^u \frac{F_{\ell}[\bar V(w)]}{\sqrt{1 - \bar V(w)}} \dd{w} \, ,\\
  \scS_3(u) &=-\frac{1}{24 } F_{\ell}[\bar V(u)]  \, ,
\end{align}
where
\begin{equation}
  F_{\ell}[\bar V(u)] = \bar V''(u) + \frac{3}{u} \bar V'(u) - \frac{2 \ell (\ell + 1)}{u^2} \pqty*{ 1-\bar V(u) } \, .
\end{equation}
At \ac{nlo} we see the contribution of the angular momentum \(\ell\) that in the physical optics approximation did not change the energy and only accounted for a large degeneracy.
While here the degeneracy is lifted, the \(1/u^2\) behavior of the new term proportional to \(\ell ( \ell + 1)\) is problematic.
For \(u \to 0\) it becomes so large that it breaks the \ac{wkb} approximation and leads to an apparent singularity. 
At this order, we can estimate the region of validity of the expansion by requiring the \ac{nlo} term to be smaller than the leading term and find that for $\ell>0$ our approximation is valid for distances larger than order $\delta$ from the origin.
We expect the divergences to disappear when all orders of the expansion are resummed. In the case of the harmonic potential, treated in Section~\ref{sec:fluctuations-harmonic-potential}, we have an exact solution which does not display any singularities. Here we will concentrate on the s-wave approximation; in other words we only consider the lower rotational-invariant modes (\(\ell = 0\)).

Now we can repeat the same computation that we had performed in the previous section, fixing the integration constants and the allowed values for \(q_0\) to avoid singular solutions.
At this order we find
\begin{equation}
  \scS_1(u) + \frac{\varpi^2}{q_0^2} \scS_3(u) = - \log(u) - \frac{1}{2} \log*( 1- \bar V(u)) - \frac{\varpi^2}{24 q_0^2} F_0[\bar V(u)] + \order*{\frac{\varpi^4}{q_0^4}} \, ,
\end{equation}
that we can rewrite as
\begin{equation}
  \scS_1(u) + \frac{\varpi^2}{q_0^2} \scS_3(u) = - \frac{1}{2} \log*(u^2 + \frac{u^2 \varpi^2}{12 q_0^2} F_0[\bar V(u)]) - \frac{1}{2} \log( 1- \bar V(u)) + \order*{\frac{\varpi^4}{q_0^4}} \, .
\end{equation}
\begin{itemize}
\item Imposing regularity at \(u = 0\), we find that the solution must take the form
\begin{equation}
  \pi(u)   = D \frac{\sin*( \frac{q_0}{\varpi} \sqrt{3} \int_0^u \frac{\dd{w}}{\sqrt{1 - \bar V(w)}} + \frac{\varpi}{q_0} \frac{1}{4\sqrt{3}} \int_0^u \frac{F_{\ell}[\bar V(w)]}{\sqrt{1 - \bar V(w)} } \dd{w}  )}{u\sqrt{\pqty*{1 + \frac{\varpi^2}{12 q_0^2}F_0(\bar V(u)) } \pqty*{1 - \bar V(u)}}}  \,.
\end{equation}
\item Now we impose regularity in \(u = 1\) and find the quantization condition
\begin{align}
	\frac{q_0}{\varpi} \sqrt{3} \int_0^1 \frac{\dd{w}}{\sqrt{1 - \bar V(w)}} + \frac{\varpi}{q_0} \frac{1}{4 \sqrt{3}} \int_0^1 \frac{F_0[\bar V(w)]}{\sqrt{1 - \bar V(w)} } \dd{w} &= n \pi\, , & n&=1,2,3,\dots
\end{align}\end{itemize}

\begin{figure}
  \centering
  \includegraphics[width=.7\textwidth]{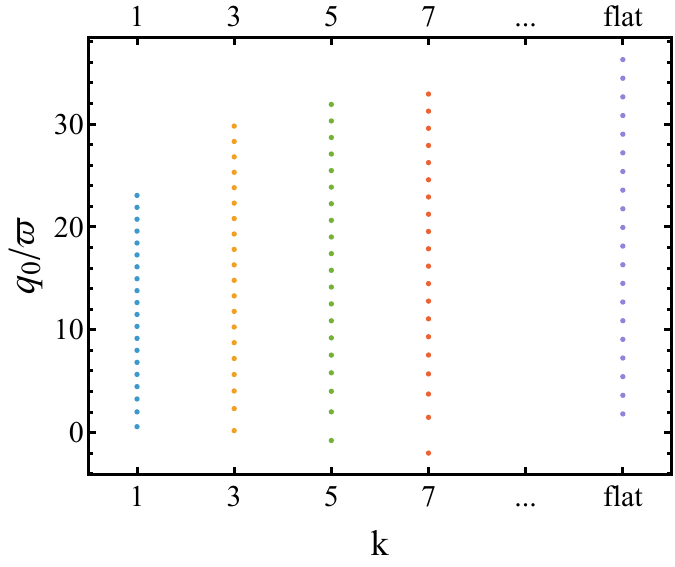}
  \caption{First twenty allowed energy eigenvalues in the spectrum at low energy for different choices of the potential \(\bar V = u^{2k}\). The eigenvalues are less dense for flatter potentials, and for \(k \to \infty \) asymptote to the spectrum of the spherical box \(q_0/\varpi = \pi n /\sqrt{3}\).}
  \label{fig:low-energy-spectrum}
\end{figure}

For a power-law potential \(\bar V  = u^{2k}\) it is possible to solve the integrals analytically to find
\begin{equation}
  \frac{q_0(n)}{\varpi} = \frac{2}{\sqrt{3} \binom{1/(2k)}{1/2}} n - \frac{2 (k+1)  }{2 \sqrt{3} \pi \binom{(k-1)/2k}{1/2}} \frac{1}{n} +\dots
\end{equation}
In the case of the harmonic potential \(k = 1\), this reduces to 
\begin{equation}
  \frac{q_0(n)}{\omega} = \frac{2}{\sqrt{3}} n  - \frac{4}{4 \sqrt{3}} \frac{1}{n} +\dots
\end{equation}
The result for some values of \(k\) is sketched in Fig.~\ref{fig:low-energy-spectrum}.
For larger values of \(k\) (flatter potentials) the spectrum becomes less and less dense, and for \(k \to \infty \) it asymptotes the spectrum of the spherical box in Eq.~\eqref{eq:spherical-box-spectrum}: \(q_0/\varpi = \pi n /\sqrt{3}\).

In the next section we will study in detail the case of the harmonic trap, that can be solved exactly in the \ac{lo} \ac{eft}, and recover this result in the appropriate limit.

% in agreement with the exact solution that we will see in Eq.~\eqref{eq:harmonic-dispersion} in the next section.
% \todo[inline]{%
% The expansion of the exact result at large $n$ ($\delta\sim n^{-1}$) is
% \[
% \frac{q_0}{\omega} =\sqrt{\tfrac{4n}{3}(n+l+2)+l}
% \xrightarrow[n\to\infty]{}
% \frac{2n}{\sqrt{3}}+\frac{l+2}{\sqrt{3}}
% -\frac{l^2+l+4}{4\sqrt{3}\,n}
% +O\!\left(\frac{1}{n^2}\right).
% \]
% There is a (different) constant term. In the case $l=0$ we can match the two results by shifting $n\longrightarrow n+1$ in WKB or expanding the exact result after shifting $n\longrightarrow  n-1$, since $n$ is large, sounds reasonable. For $l\ne 0$, we can shift n by a $l$-dependant term to match the expansion of the exact result
% }

\subsection{Harmonic Potential}
\label{sec:fluctuations-harmonic-potential}
{}\reversemarginpar\marginnote{\usebox{\lowenergyarrow}}[-1cm]%

In the low-energy regime, where the controlling parameter \(\eta\) is very small, and it is sufficient to use the \ac{lo} \ac{eft}, there is a special choice for the potential that allows us to solve the \ac{eom} exactly: 
the harmonic potential
\begin{equation}
  V = \frac{\omega^2}{2} r^2 \, .  
\end{equation}
In this case, the low energy parameter is \(\varpi = \omega\). For this section, we will keep the ratio \(\delta = \varpi/q_0\) finite.

The \ac{eom} for the radial mode is
\begin{equation}
  \delta^2 \odv{}{u} \pqty*{u^2 \pqty{1 - u^2}^{3/2} \odv{}{u} \pi(u)} - \delta^2 \ell (\ell + 1) \pqty{ 1 - u^2}^{3/2} \pi(u) + 3 u^2 \pqty*{1 - u^2}^{1/2} \pi(u) = 0 \, .
\end{equation}
As already observed in~\cite{Kravec:2018qnu}, after the rescaling \(\pi(u) = u^{\ell} w(u) \) and the change of variable \(u^2 = z\), the \ac{eom} takes the form of a hypergeometric equation:
\begin{equation}
  z \pqty*{1 - z} \odv[order=2]{w}{z} + \pqty*{ \frac{3}{2} + \ell - \pqty{3 + \ell } z } \odv{w}{z} - \frac{3}{4} \pqty*{ \ell - \frac{1}{\delta^2}  } w = 0 .
\end{equation}
Its solution is \(\pFq{2}{1}{a,b}{c}{z}\) with parameters
\begin{align}
   a &=\frac{\ell +2 - \sqrt{\ell ^2+\ell +4 +3/\delta^2}}{2 }, & b&= \frac{\ell +2+\sqrt{\ell ^2+\ell +4 +3/\delta^2}}{2 }, & c&= \frac{3}{2} + \ell \, .
\end{align}
We want a regular solution in \(z = 1\).
The general theory of Gaussian hypergeometric functions requires that, if \(a + b > c\), \(a\) must be a non-positive integer (\emph{i.e.}, \(a= - n\), where \(n = 0, 1, \dots\)).
This condition is precisely what causes the hypergeometric function to degenerate into a polynomial, and it directly leads to the expected quantization condition for \(\delta\):
\begin{align}
  \delta^2 &= \frac{3}{4 n^2+4 n \ell +8 n+3 \ell } , & n, \ell &= 0, 1, \dots \, . \, .
\end{align}
The final result is that the fluctuations are written as 
\begin{equation}
  \pi(t, \mathbf{u}) = e^{i q_0 t} \textstyle \pFq{2}{1}{-n, n + \ell+ 2}{3/2 + \ell}{u^2} Y_{\ell m}(\vartheta, \phi)
\end{equation}
with the spectrum
\begin{align}
  q_0(n,l) &= \omega \sqrt{\frac{4n}{3}\pqty*{n + \ell + 2} + \ell} , & n, \ell &= 0, 1, \dots \, .
\end{align}

In the special case of \(\ell = 0\), the s-wave approximation on which we will concentrate also in the following, the radial part is a Chebyshev polynomial of the second kind:
\begin{equation}
  \pi^{n,0}_r(u) =\textstyle \pFq{2}{1}{-n, n + 2}{3/2}{u^2} = U_n( 1 - 2 u^2)\,,
\end{equation}
which can be written in terms of trigonometric functions as
\begin{equation}
  \pi^{n,0}_r(u) = U_n( 1 - 2 u^2) =  \frac{ 2 
  \sin*( 2(n+1) \arcsin(u) )}{u \sqrt{1- u^2}} \, .
\end{equation}
This expression is precisely the same as the one found in Eq.~\eqref{eq:physical-optics-harmonic-oscillator}, up to the shift \(n \to n + 1\).
In the special case of the harmonic oscillator, the functional form obtained using the physical optics \ac{wkb} approximation turns out to be exact.
The quantization condition shows that \(\delta\) becomes small when \(n\) is large, allowing us to approximate \(n\) as \(n+1\) in this limit.

\section{The dynamic structure factor}
\label{sec:dynamic-structure-factor}

The phonon field and its fluctuations are not observable quantities, and therefore it is imperative to identify an observable which encodes the acoustic excitation spectrum
of the superfluid. In the language of field theory, observables are matrix elements of local operators. Here, however, we are in the non-relativistic regime and therefore the probability
to excite the superfluid many-body system away from the ground state is, via Fermi's golden rule, proportional to the so-called
dynamic structure factor~\cite{Sturm+1993+233+242}
\begin{equation}
  \label{eq:S-Fermi-golden}
  S(q_0, \mathbf{q}) = \sum_{\mathbf{n}} \abs{ \braket{0 | \delta \rho(\mathbf{q})| \mathbf{n}}}^2 \delta(q_0 - q_0(\mathbf{n})),
\end{equation}
where \(\delta \rho(\mathbf{q})\) is the Fourier transform of the
small fluctuations induced in the charge density about the ground
state, \(\mathbf{n}\) is the set of quantum numbers that identify
the corresponding quantum states and \(q_0(\mathbf{n})\) is the spectrum equation\footnote{For a review, see Ref.~\cite{Sturm+1993+233+242}. In an appendix we provide a field-theoretic definition of the dynamic structure factor and response function.}.

\subsection{LO EFT}
\label{sec:S-LO-EFT}
%{}\reversemarginpar\marginnote{\usebox{\highenergyarrow}}[-1cm]%

For ease of exposition we start from the \ac{lo} \ac{eft}.
It is convenient to rescale the field \(\pi(t, \mathbf{u})\) so that its kinetic term is canonically normalized, which in this case means that the coefficient of the time derivative part of the Lagrangian is \(1/2\).
The rescaled field \(\tilde \pi(t, \mathbf{u})\) is then
\begin{equation}
  \tilde \pi(t, \mathbf{u}) = \mu^{1/4}\frac{\sqrt{15 c_0}}{2} \pqty*{ 1 - \bar V(r)}^{1/4} \pi(t, \mathbf{u}) \, ,
\end{equation}
where \(\pi(t, \mathbf{u})\) is the solution that we have found in various approximations in the previous section.
The charge density is the variation of the action with respect to \(\dot \pi\), so in terms of the fluctuations \(\tilde \pi(t, \mathbf{u})\), the leading order fluctuations of the charge are written as
\begin{equation}
   \delta \rho = \frac{\sqrt{15 c_0}}{2} \mu^{1/4} \pqty*{ 1 - \bar V}^{1/4} \pdv{}{t} \tilde \pi(t, \mathbf{u})  \, .
\end{equation}
  
Now we need to canonically quantize starting with the field \(\tilde \pi(t, \mathbf{r})\).
The first step consists in identifying a basis of solutions for the \ac{eom}.
We have performed a position-dependent rescaling from \(\pi(t, \mathbf{q})\): this changes the form of the Sturm--Liouville problem (Section~\ref{sec:linear-regime}) and the weight function in the associated scalar product.
The (non-geometric part of the) weight is just the coefficient of the second time derivative in the \ac{eom}, which we have now rescaled to one.
It follows that the \(\tilde \pi(\mathbf{u})\) are orthogonal with respect to the usual flat-space product:
\begin{equation}
  \braket{ \tilde \pi^{nlm}, \tilde \pi^{n' l' m'}} = \int_0^1 u^2 \dd{u} \int_{S^2} \dd{\Omega} \tilde \pi^{n\ell m}(u, \theta, \phi ) \tilde \pi^{n' \ell' m'}(u, \theta, \phi ) = \delta_{nn'} \delta_{\ell \ell'} \delta_{mm'} \, .
\end{equation}
To add the time component, we observe that the system is invariant under time translation. We can thus use the standard
equal-time product associated with the \ac{eom}:
\begin{equation}
  ( \phi_1(t, \mathbf{u}), \phi_2(t, \mathbf{u}) ) = i \int_{u\le1} \dd^3{\mathbf{u}}  \phi_1(t,\mathbf{u})^*\overset{\leftrightarrow}{\partial_t} \phi_2(t,\mathbf{u}) \, ,
\end{equation}
where $f\overset{\leftrightarrow}{\partial_t} g\equiv f(\partial_t g)-(\partial_t f)g$. With this we can identify a basis of orthonormal solutions.
In the physical optics approximation they are
\begin{equation}
  \phi_{n \ell m} (t, \mathbf{u}) = \frac{D}{2\sqrt{q_0}} e^{-i q_0 t} \frac{\sin(\scS_0(u)/\delta)}{u \pqty*{ 1- \bar V}^{1/4}} Y_{\ell m}(\theta, \phi) \, ,
\end{equation}
where the \(q_0\) are quantized as in Eq.~\eqref{eq:q0-quantization}
\begin{align}
   \frac{q_0}{\varpi} \scS_0(1) = n \pi, && n \in \setZ \, ,
\end{align}
and \(D\) is the normalization constant in Eq.~\eqref{eq:fluctuation-normalization}.
From here on, the procedure is standard.
We introduce standard creator and annihilator operators and a Fock space starting from the vacuum \(\ket{0}\) corresponding to the ground state
\begin{align}
  [ a_{n \ell m}, a^{\dagger}_{n' \ell' m'}] &= \delta_{n n'} \delta_{\ell \ell'} \delta_{m m'} \, , & \ket{n \ell m} = a^{\dagger}_{n \ell m} \ket{0} \, .
\end{align}
The quantized field reads
\begin{equation}
  \hat \pi^{n \ell m}(t, \mathbf{u}) = \sum_{n \ell m} \phi_{n \ell m}(t, \mathbf{u}) a_{n \ell m} + \phi_{n \ell m}^{*}(t, \mathbf{u}) a^{\dagger}_{n \ell m} \, .
\end{equation}
With this we can read off the small fluctuations:
\begin{equation}
  \delta \hat \rho(t, \mathbf{u}) = \sum_{n \ell m} e^{-i q_0 t} F_{n \ell m}(\mathbf{u}) a_{n \ell m} + e^{i q_0 t}  F_{n \ell m}^{*}(t, \mathbf{u}) a^{\dagger}_{n \ell m} ,
\end{equation}
where
\begin{equation}
 F_{n \ell m}(\mathbf{u})  = \frac{i}{2} \mu^{1/4} \sqrt{\frac{15 c_0 q_0}{Y}} \frac{\sin(\scS_0(u)/\delta)}{u} Y_{\ell m}(\theta, \phi)
\end{equation}
with the normalizations chosen so that
\begin{equation}
    \braket{0 | \delta \rho(\mathbf{q})| n \ell m} = i \mu^{1/4} D \sqrt{ 15 c_0 q_0} \int \frac{\dd^3{\mathbf{u}}}{(2\pi)^3} e^{i R_{cl} \mathbf{q} \cdot \mathbf{u}}  \frac{\sin(\scS_0(u)/\delta)}{u} Y_{\ell m}(\theta, \phi) \, .
\end{equation}

\begin{figure}
  \centering
  \begin{tikzpicture}
    % Include the image (width can be adjusted)
    \node[anchor=south west, inner sep=0] (img) at (0,0)
    {\includegraphics[width=0.47\textwidth]{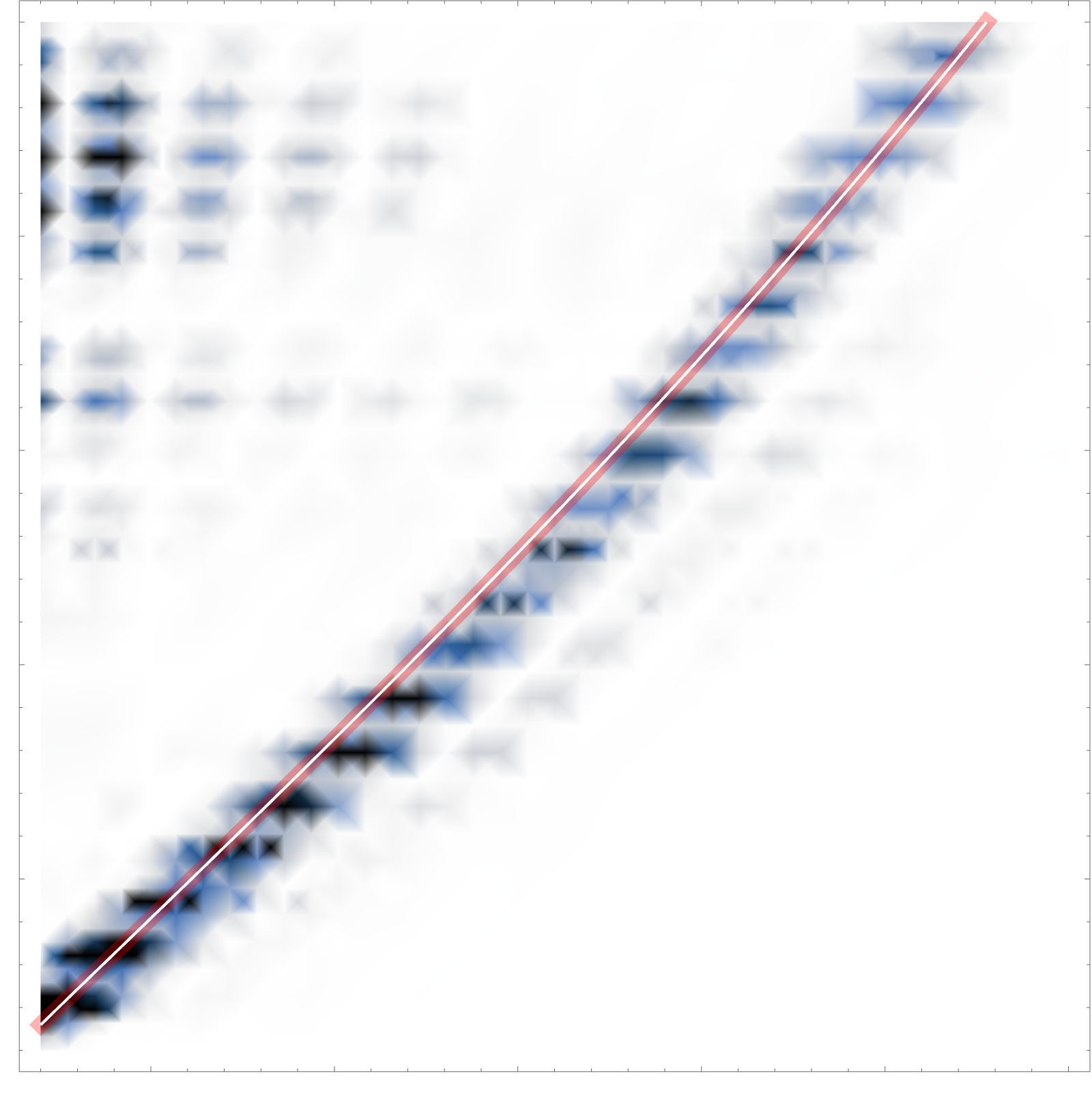} \includegraphics[width=0.47\textwidth]{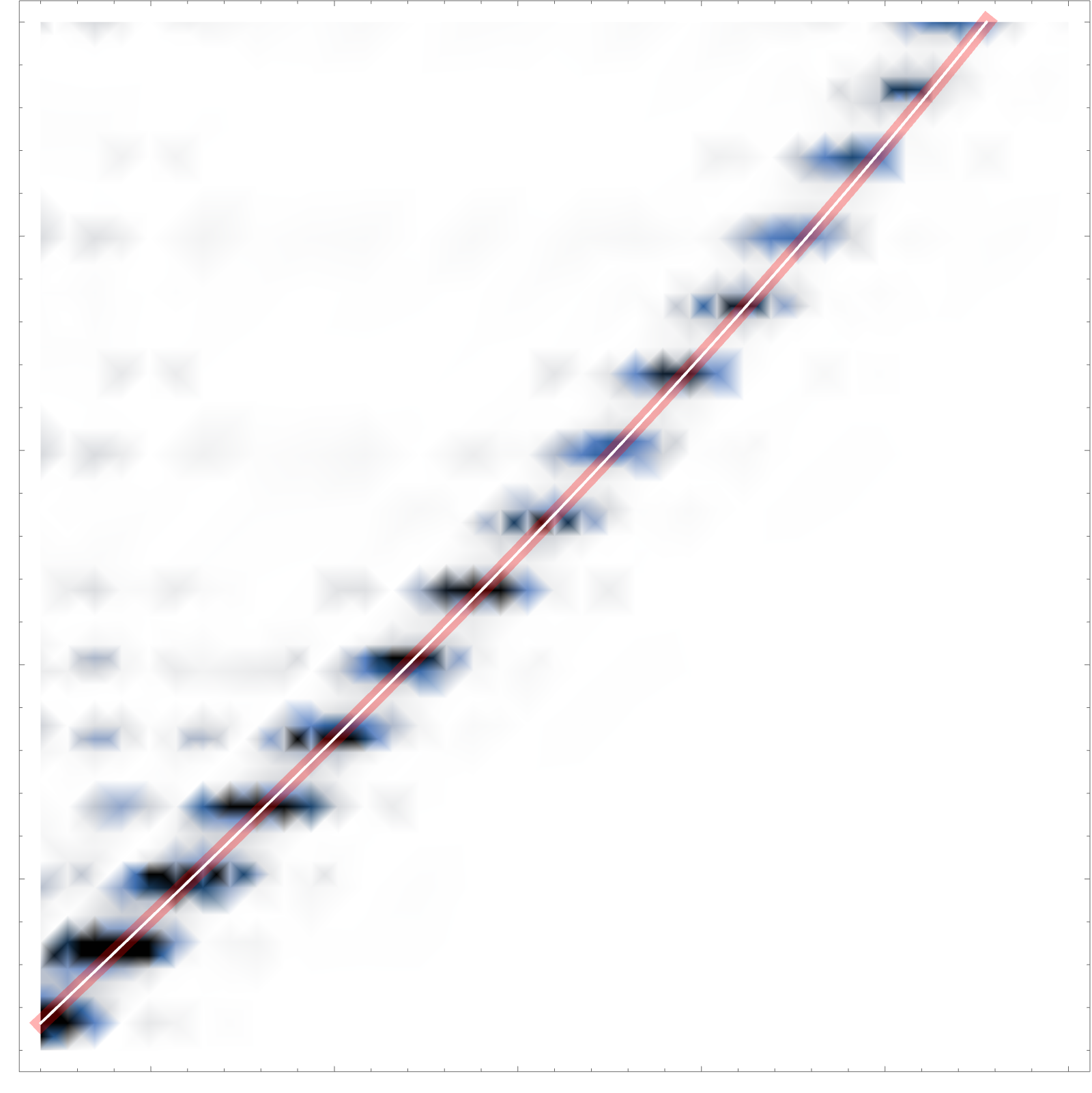}};

    % Make coordinates relative to the image
    \begin{scope}[x={(img.south east)}, y={(img.north west)}]
      \foreach \y in {0.1,0.2,...,0.9}
      {
        \draw (0.01,\y) -- (0.015,\y);
        \draw (0.9975,\y) -- (.99,\y);
      }

      \foreach \x in {0.1,0.2,...,0.9}
      {
        \draw (\x,0.02) -- (\x, 0.035);
%        \draw (0.9975,\y) -- (1.01,\y);
      }

      \shade[bottom color=red, top color=blue] (.49,0.02) rectangle (.515,1);
      \node[opacity=.8] at (.425,0.1) {\footnotesize low energy};
      \node[opacity=.8] at (.59,0.55) {\footnotesize linear regime};
      \node[opacity=.8] at (.585,0.9) {\footnotesize high energy};

      \draw [draw=black] (.01,.02) rectangle (.997,.997);

      \node[opacity=.8] at (.25,-0.05) { \(q/\sqrt{\varpi}\)};
      \node[opacity=.8] at (.75,-0.05) { \(q/\sqrt{\varpi}\)};

      \node[opacity=.8] at (.515,-.01) {\scriptsize \(0\)};
      \node[opacity=.8] at (.995,-.01) {\scriptsize \(3\)};

      \node[opacity=.8] at (.015,-.01) {\scriptsize \(0\)};
      \node[opacity=.8] at (.49,-.01) {\scriptsize \(3\)};

      \node[opacity=.8, rotate=90] at (-.02,.5) { \(q_0 / \varpi\)};
      \node[opacity=.8] at (-.01,.995) {\scriptsize \(25\)};
      \node[opacity=.8] at (-.01,0.02) {\scriptsize \(1\)};

      \node[opacity=.8, rotate=90] at (1.02,.5) { \(q_0 / \varpi\)};
      \node[opacity=.8] at (1.025,.995) {\scriptsize \(25\)};
      \node[opacity=.8] at (1.025,0.02) {\scriptsize \(1\)};

    \end{scope}
  \end{tikzpicture}

  \vspace{2em}
  
  \begin{tikzpicture}
    % Include the image (width can be adjusted)
    \node[anchor=south west, inner sep=0] (img) at (0,0)
    {\includegraphics[width=0.47\textwidth]{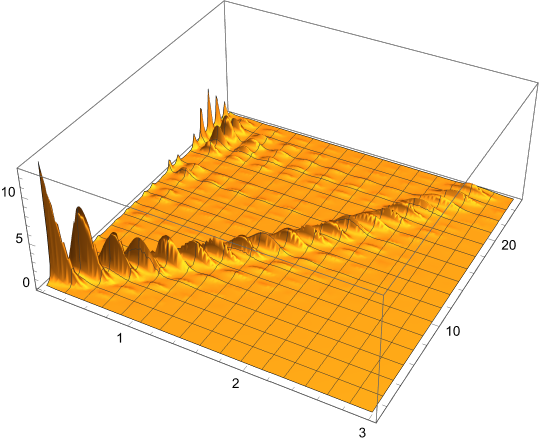} \includegraphics[width=0.47\textwidth]{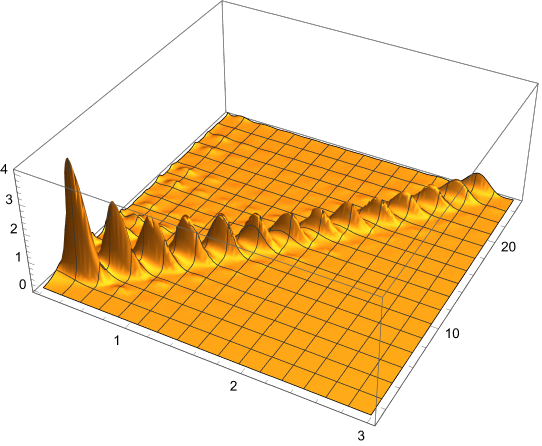}};

    % Make coordinates relative to the image
    \begin{scope}[x={(img.south east)}, y={(img.north west)}]

      % \shade[bottom color=red, top color=blue] (.49,0.02) rectangle (.515,1);
      % \node[opacity=.8] at (.425,0.1) {\footnotesize low energy};
      % \node[opacity=.8] at (.59,0.55) {\footnotesize linear regime};
      % \node[opacity=.8] at (.585,0.9) {\footnotesize high energy};

      % \draw [draw=black] (.01,.02) rectangle (.997,.997);

      \node[opacity=.8] at (.15,0) {\footnotesize \(q/\sqrt{\varpi}\)};
      \node[opacity=.8] at (.65,0) {\footnotesize \(q/\sqrt{\varpi}\)};

      \node[opacity=.8] at (.425,.2) {\footnotesize \(q_0 / \varpi\)};
      \node[opacity=.8] at (.925,.2) {\footnotesize \(q_0 / \varpi\)};

      % \node[opacity=.8] at (.515,-.01) {\scriptsize \(0\)};
      % \node[opacity=.8] at (.995,-.01) {\scriptsize \(3\)};

      % \node[opacity=.8] at (.015,-.01) {\scriptsize \(0\)};
      % \node[opacity=.8] at (.49,-.01) {\scriptsize \(3\)};

      % \node[opacity=.8, rotate=90] at (-.02,.5) { \(q_0 / \varpi\)};
      % \node[opacity=.8] at (-.01,.995) {\scriptsize \(25\)};
      % \node[opacity=.8] at (-.01,0.05) {\scriptsize \(1\)};

    \end{scope}
  \end{tikzpicture}
  \caption{
    Dynamic structure factor for a harmonic trap (left) and a superharmonic confining potential \(\bar V = u^{16}\) (right) for \(\mu / \varpi = 100\).
    The continuous lines show the curves along which \(S(q_0,q)\) is peaked (dispersion relations).
    The discrete structure reflects the finite size of the system.
    The width of the peaks in the \(q\) direction results from the breaking of translational invariance.
    Note that the peaks are more sharply defined in the superharmonic case.
    A Lorentzian smoothing with parameter \(\Delta q_0 = .5 \varpi\) has been applied in the \(q_0\) direction since the system does not break time-translation invariance.
  }
  \label{fig:S-harmonic}
\end{figure}
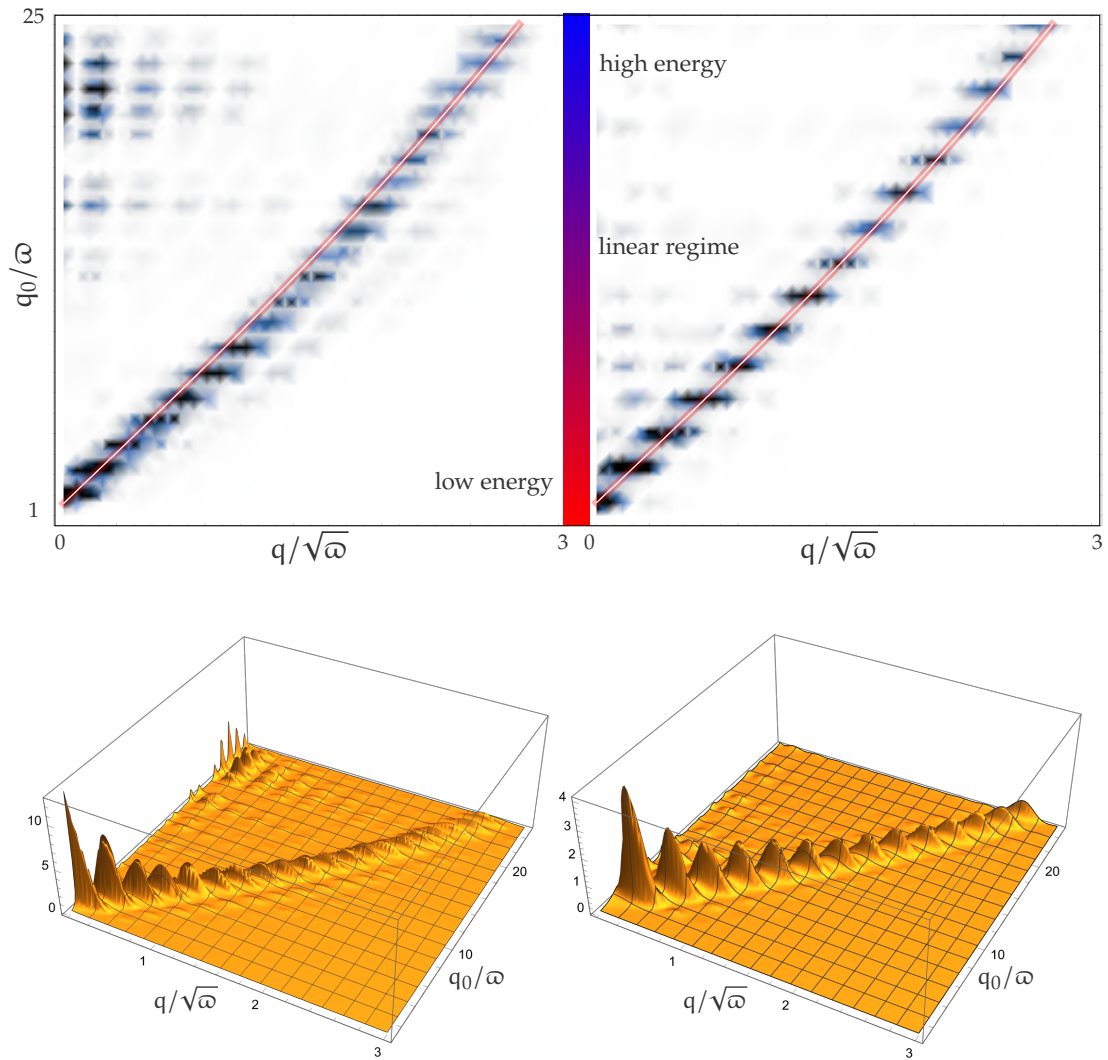

The problem has radial symmetry and the angular part of the integral boils down to the decomposition of a plane wave in terms of spherical harmonics
\begin{equation}
  \int \dd{\Omega} Y_{\ell m} e^{i\mathbf{q}\cdot\mathbf{r}} = 4 \pi \delta_{\ell \ell'}\delta_{m m'}  i^{\ell'}j_{\ell'} (qr) Y_{l'm'}\,, 
\end{equation}
where the \(j_{\ell}\) are Bessel functions.
In the special case of $\ell =0$, this is simply 
\begin{equation}
   \int \dd{\Omega}  Y_{0 0} e^{i\mathbf{q}\cdot\mathbf{r}} = 4 \pi j_0(qr) Y_{00} = 2 \sqrt{\pi} \frac{\sin(qr)}{qr} \, ,
\end{equation}
so, in the s-wave approximation, the matrix element of the charge fluctuation operator  is the Fourier sine transform of the phase of the fluctuations:
\begin{equation}\label{eq:vevofDensityFluc}
   \braket{0 | \delta \rho(\mathbf{q})| n } \propto \frac{\mu^{1/4} \sqrt{c_0 q_0}}{q} \int_0^1 \dd{u} \sin(q R_{cl} u) \sin*(\frac{\scS_0(u)}{\delta}) \, .
\end{equation}
This expression relates the study of the fluctuations of the previous section directly to an experimentally-accessible quantity.
Unfortunately, the integral cannot be solved analytically in general.
One possible way out, then, is to perform a numerical analysis (see Fig.~\ref{fig:S-harmonic}).
Alternatively, as we will do in the following section, one can look for approximate solutions that, in different energy regimes are controlled by the small parameter \(\eta\) or \(\delta\).
These approximate solutions allow us to define a notion of dispersion relation even in the absence of translation invariance and to discuss the effects of both a non-trivial potential and of higher-order terms in the \ac{eft}.

% \end{comment}
%%%%%%%%%%%%%%%%%%%%%%%%%%%%%%%%%%%%%%%%%%%%%%%%%%%%%
\subsection{High energy regime}
\label{sec:high-energy-regime-S}
{}\reversemarginpar\marginnote{\usebox{\highenergyarrow}}[-1cm]%

If \(q R_{cl} \gg 1\) with \(q R_{cl} = \order{1/\delta}\), the main contribution to the integral comes from the saddle.
Since \(q >0\) and \(u>0\), for \(\scS_0' > 0\), this is the value \(u = \bar u\) such that
\begin{equation}
  q R_{cl}= \frac{1}{\delta} \scS_0'(\bar u)
\end{equation}
and the integral is well approximated by
\begin{equation}\label{eq:int-approx}
  \braket{0 | \delta \rho(\mathbf{q})| n \ell m}  \approx %\frac{\mu^{1/4} \sqrt{c_0 q_0}}{q} \sqrt{\frac{2 \pi}{\abs{S''(\bar u) \delta}}} \sin( q R_{cl} \bar u) e^{i/\delta \scS_0(\bar u)}
  \mu^{1/4} \sqrt{\frac{2 \pi c_0}{\varpi}} \frac{q_0}{q} \frac{1}{\sqrt{\abs{\scS''(\bar u)}}} \sin( q R_{cl} \bar u)  \sin*(\frac{\scS_0(u)}{\delta}).
\end{equation}
At this order, the dynamic structure factor becomes
\begin{equation}
  S(q_0, \mathbf{q}) \approx \mu^{1/2} \frac{c_0}{\varpi q^2} \eval*{\sum_n \frac{q_0^2}{\abs{\scS_0''(\bar u)}} \sin^2(q R_{cl} \bar u) \delta(q_0 - \tfrac{n \pi \varpi}{\scS_0(1)}) }_{q = \frac{q_0}{\sqrt{2\mu}} \scS_0'(\bar u)} \, .
\end{equation}
As expected we find a discrete sum of oscillating functions.
We are interested in the position of the peaks of \(S(q_0, \mathbf{q})\).
Up to subleading correction in \(1/\delta\), these functions in the sum are peaked for the values of \((q_0, q)\) such that \(\scS_0''(\bar u(q_0, q)) = 0\).

\paragraph{Linear regime.}
In the \emph{linear regime}, where it is sufficient to study the physical optics approximation of the \ac{lo} \ac{eft}, we have found that \(\scS_0\) is given by Eq.~\eqref{eq:physical-optics-fluctuations}:
\begin{equation}
  \scS_0(u) = \sqrt{3} \int_0^u \frac{\dd{w}}{\sqrt{1 - \bar V(w)}},
\end{equation}
so the saddle-point equation is
\begin{equation}
  q = \frac{q_0}{ \sqrt{2 \mu}} \frac{\sqrt{3}}{\sqrt{1 - \bar V(u)}} = \sqrt{\frac{3}{2 \pqty*{\mu - V(r)}}} q_0 \, .
\end{equation}
This equation is reminiscent of a \acl{lda} for the dispersion relation in a slowly-varying potential (remember that in vacuum \(q_0 = \sqrt{2 \mu/3}q\)).
However, one should not read it as a position-dependent speed of sound, as it is simply the curve $u= u(q_0, \mathbf{q})$ on which the Fourier integral in Eq.~\eqref{eq:vevofDensityFluc} localizes. 

In a system with translation invariance, the Fourier integral would yield delta functions and the dynamic structure factor would be completely localized along the curve of the dispersion relation.
The potential breaks translational invariance, so $S$ is a more complicated function.
It is natural to define an approximate dispersion relation as the curve in the \((q_0,q)\) plane along which the dynamic structure factor is peaked.
To see where this happens, we need to find the zeros of \(\scS_0''(\bar u)\):
\begin{equation}
  \label{eq:LO-peaks}
  \scS_0''(\bar u) = - \frac{\sqrt{3} \bar V'(\bar u)}{2\pqty*{ 1 - \bar V(\bar u)}^{3/2}} = 0\, .
\end{equation}
So we have peaks corresponding to the stationary points of \(\bar V(u)\). For a power-law potential, this means \(\bar u = 0\).
% one needs to solve for \(u\)
% \begin{equation}
%    \bar u = \bar V^{-1} \pqty*{1 - \tfrac{3 q_0^2}{2 \mu q^2}} \, ,  
% \end{equation}
% and then maximize \emph{w.r.t.} \(q\).
The result is that, at this order, the dynamic structure factor is peaked for
\begin{equation}\label{eq:peakDSF}
  q = \eval*{ \sqrt{\frac{3}{2 \pqty*{\mu - V(r)}}} q_0}_{r = 0} = \sqrt{\frac{3}{2 \mu}} q_0 ,
\end{equation}
which is \emph{independent of the potential} (that vanishes at the origin).
As expected, the leading behavior of the spectrum is linear, since the fluctuations around the ground state are described by a type-I Goldstone boson~\cite{Son:2005rv,Favrod:2018xov}.

\bigskip

The width of the peak \(\Delta q\) will depend on the potential.
To estimate it, start from the approximate expression for the matrix element of the density fluctuations in Eq.\eqref{eq:int-approx}.
We can just approximate the width of the peak in the dynamic structure factor with the position of the first zero in the sin function
\begin{equation}
  \Delta q \Delta u R_{cl} = \pi \, .
\end{equation}
For a power-law potential \(V = u^{2k}\), the variations \(\Delta q\) and \(\Delta u\) are related by
\begin{equation}
  (\Delta u)^{2k} \approx \Delta q R_{cl} \delta  
\end{equation}
and we find that
\begin{equation}
   \Delta q = \order*{\frac{1}{R_{cl}} \pqty*{ \frac{q_0 R_{cl}}{\sqrt{\mu}}}^{1/(1+2k)} }.
\end{equation}
The larger \(k\), the flatter the potential and the narrower the peak.

\paragraph{\ac{nlo} in the \ac{eft}.}
The construction generalizes to higher orders in the \ac{eft}.
If we go to \ac{nlo} in the \ac{eft} and remain at \ac{lo} in \ac{wkb}, consistently with the limit
\begin{equation}
  \frac{\varpi}{q_0} < \frac{q_0}{\mu} \ll 1 \, ,
\end{equation}
there are two main effects to take into account:
\begin{enumerate}
\item The normalization of the fluctuations now becomes \(q\)-dependent, reflecting the fact that the \ac{eom} is of higher order in derivatives;
\item The function \(\scS_0(u)\) now contains a higher-order term in the \(q_0/\mu\) expansion (see Eq.~\eqref{eq:NLO-EFT-LO-WKB}).
\end{enumerate}
Here we will limit ourselves to studying how the \ac{nlo} corrections in the \ac{eft} affect the dispersion relation.
A detailed analysis of the behavior of the dynamic structure factor at \ac{nlo} requires a different approach based on the path-integral formulation of this problem as outlined in Appendix~\ref{sec:resp-funct-path-integral}.

The saddle-point equation now becomes 
\begin{equation}
  q = \frac{q_0}{\sqrt{2\mu}} \scS_0'(0) = \sqrt{ \frac{3}{2\mu} } q_0\pqty*{1+\frac{2}{5c_0}(c_1-3c_2)\frac{q_0^2}{\mu^2} + \order*{\frac{q_0}{\mu}}^4},
\end{equation}
and the dynamic structure factor is peaked again for \(\bar u = 0\), \emph{i.e.} along the curve
\begin{equation}
  q_0(q)= \sqrt{\frac{2 \mu}{3}} q \pqty*{ 1 - \frac{4}{15} \frac{c_1 - 3 c_2}{c_0} \frac{q^2}{\mu}}  ,
\end{equation}
consistent with Eq.~\eqref{eq:no-trap-dispersion}.
In the high-energy regime that we are discussing here, the control parameter is \(\eta = q_0/\mu\), which depends only on the energy of the fluctuations and on the chemical potential.
It is not surprising, then, that the leading correction to the linear behavior of the dispersion relation in this region is independent of the details of the trapping potential and is in this sense universal. With the mean-field value of the low energy constant given in Section \ref{sec:sfeft}, the curvature is always convex.
Other aspects of the dynamic structure factor (for example the width of the peak) do depend on \(V\), but this goes beyond the scope of this work.
The situation is quite different in the low-energy regime as we will see in the following section.

%%%%%%%%%%%%%%%%%%%%%%%%%%%%%%%%%%%%%%%%%%%%%%%%%%%%%
\subsection{Low energy regime}
\label{sec:low-energy-regime-S}
{}\reversemarginpar\marginnote{\usebox{\lowenergyarrow}}[-1cm]%

There are two sources of corrections to the position of the peaks of the dynamical structure factor in the low-energy regime ($\delta \ll 1$).
The \ac{nlo} term $\scS_2$ in the \ac{wkb} expansion of the wavefunction that we have discussed in Sec.~\ref{sec:low-energy-regime} and the expansion of the integral in Eq.~\eqref{eq:int-approx} to \ac{nlo} beyond the saddle-point approximation. The quantity to compute is 
\begin{equation}
	I = \int_0^1 \dd{u} \exp*[i\pqty*{qR_{cl} u - \frac{\scS_0(u)}{\delta} - \scS_2(u) \delta }].
\end{equation} 
We know that the most important contribution comes from the saddle. For ease of exposition, we introduce
\begin{align}
	\xi &=q R_{cl} \delta, & \phi_\xi(u) &= \xi u -\scS_0(u) - \scS_2(u) \delta^2 \, ,
\end{align}
so that the integral takes the form
\begin{equation}
	I =  \int_0^1 \dd{u} e^{i/\delta \phi_\xi(u)} \, ,
\end{equation}
which we expand at \ac{nlo} around the saddle $u = u_\xi$ fixed by
\begin{equation}
	\phi'_\xi(u) = \xi - \scS_0'(u_\xi) - \scS_2'(u_\xi) \delta^2 = 0.
\end{equation}
The integral expanded at quartic order in $\hat u = u-u_\xi$ is given by
\begin{multline}
	I \approx \int_{-\infty}^{\infty} \dd{\hat u} \exp*[\frac{i}{\delta}\pqty*{\phi_\xi (u_\xi) + \frac{\phi_\xi''(u_\xi)}{2}\hat u^2 + \frac{\phi_\xi^{(4)}(u_\xi)}{4!}\hat u^4}] \\
	\approx \sqrt{\frac{2\pi\delta}{\abs{\phi_\xi''}}}e^{\frac{i}{\delta} \phi_\xi(u_\xi) } \pqty*{1 - \frac{i\delta}{8}\frac{\phi_\xi^{(4)}(u_\xi)}{ \phi_\xi''(u_\xi)^2}}.
\end{multline}
We exponentiate the parenthesis to find
\begin{equation}
	I \approx  \sqrt{\frac{2\pi\delta}{\abs{\phi_\xi''}}} \exp*[\frac{i}{\delta}\pqty*{\phi_\xi(u_\xi)-\frac{\delta^2}{8}\frac{\phi_\xi^{(4)}(u_\xi)}{\phi_\xi''(u_\xi)^2}}] \, .
\end{equation}

Once more we are interested in the position of the peaks, which arise when the denominator vanishes.
The condition in Eq.~\eqref{eq:LO-peaks} becomes
\begin{equation}
    \phi_\xi'' = - \scS_0''(u) - \scS_2''(u) \delta^2 = 0 \, .
\end{equation}
Going back to the initial variables, the integral giving the expectation value of the density fluctuations localizes around the curve
\begin{equation}
  q = \sqrt{\frac{3}{2 \pqty{ \mu - V }}} q_0 \bqty*{ 1 + \frac{1}{6 q_0^2} \pqty*{ V''(r) + 3 \frac{V'(r)}{r}}} .
\end{equation}
The peaks of the dynamic structure factor appear at the roots of the equation
\begin{equation}
    \odv{}{r} \bqty*{ \sqrt{\frac{3}{2 \pqty{ \mu - V }}} q_0 \bqty*{ 1 + \frac{1}{6 q_0^2} \pqty*{ V''(r) + 3 \frac{V'(r)}{r}}} } = 0.
\end{equation}
For a power-law potential $r^{2k}$ the roots are still found at \(r = 0\).
Plugging this back into the saddle-point equation, we see that for \(k > 1\), at \ac{nlo} in the \ac{wkb} expansion, the position of the peaks remains unchanged with respect to the physical optics approximation.
The only exception at this order is the harmonic potential \(k =1\), where we find the dispersion relation
\begin{align}
  q_0 = \sqrt{\frac{2\mu}{3}} q \pqty*{1 - \frac{\omega^2}{q^2 \mu} + \dots} && \text{(harmonic potential)}
\end{align}
which is concave.

To determine the parametric dependence of the first non-vanishing correction for a flatter potential, we estimate how many derivatives of \(V(r)\) are required to obtain a non-zero result at \(r = 0\).
For \(V \propto r^{2k}\), this occurs at the \(2k\)-th derivative. Since \(R_{\text{cl}}\) sets the characteristic length scale, we expect a term of order \(\order{R_{\text{cl}}^{2k}}\).
Substituting \(R_{\text{cl}} = \sqrt{2\mu}/\varpi\) and using the only remaining scale, \(q\), to construct a dimensionless correction, we conclude that for a superharmonic potential the leading low-energy correction to the dispersion relation must take the form
\begin{align}
  q_0 = \sqrt{\frac{2 \mu}{3}} q \pqty*{1 + \order*{\frac{\omega^{2k}}{q^{2k} \mu^k}} }&& \text{(superharmonic potential).}
\end{align}

In the low-energy regime, the control parameter is \(\delta = \varpi/q_0\), which depends explicitly on the scale fixed by the trapping potential.
In contrast with the result of the previous section, the leading correction to the linear behavior of the dispersion relation depends explicitly on the details of \(V(r)\). 
The flatter the potential (larger values of \(k\)), the more the deviation from the linear behavior is suppressed.

%%%%%%%%%%%%%%%%%%%%%%%%%%%%%%%%%%%%%%%%%%%%%%
\subsection{Harmonic Potential}
\label{sec:S-harmonic-potential}
{}\reversemarginpar\marginnote{\usebox{\lowenergyarrow}}[-1cm]%

In Section~\ref{sec:fluctuations-harmonic-potential} we have seen that the \ac{eom} for the harmonic trap can be solved exactly in the low energy regime, and that in the s-wave limit we can write the radial part of the wavefunction as 
\begin{align}
  \pi^{n,0}_r(u) &= U_n( 1 - 2 u^2) = \frac{2 \sin*( 2(n+1) \arcsin(u) )}{u \sqrt{1 - u^2}} \, , & q_0 &= \omega \sqrt{\frac{4n}{3} \pqty*{n+2}} .
\end{align}
The low-energy regime \(\omega/q_0 \ll 1\), coincides with the large-\(n\) limit, and we can follow the steps in Section~\ref{sec:high-energy-regime-S} to estimate the value the integral in the density fluctuations with a saddle point approximation.

The integral in the expectation value for the density fluctuations localizes around the saddle at
\begin{equation}
    q = \frac{2 \pqty{n+1}}{R_{cl} \sqrt{1 - u^2}}
\end{equation}
or, in terms of \(q_0\):
\begin{equation}
q_0 = \sqrt{\frac{2 \mu \pqty{1 - u^2}}{3}\pqty*{q^2 - \frac{2 \omega^2}{\mu \pqty{1 - u^2}}}}  \, .
\end{equation}
The peak of the dynamic structure factor is at \(u = 0\), leading to the dispersion relation
\begin{equation}
  q_0 = \sqrt{\frac{2 \mu}{3}\pqty*{q^2 - \frac{2 \omega^2}{\mu}}} = \sqrt{\frac{2\mu}{3}} q \pqty*{1 - \frac{\omega^2}{q^2 \mu} + \dots},
\end{equation}
which is consistent with what we had found above at \ac{nlo} in the \ac{wkb} expansion.

% to write the function \(H(q)\) as a Legendre transform
% \begin{equation}
%   H(q) = \max_u( q R_{cl} u - 2 (n+1) \arcsin(u)) %= \eval*{q R_{cl} u_q - 2 (n+1) \arcsin(u_q)}_{u_q = } 
% \end{equation}
% so that the peak is for \(u_q = \sqrt{1 - \pqty*{\frac{2 (n+1)}{q R_{cl}}}^2} = 0\), \emph{i.e.} for
% \begin{equation}
%   q = \frac{2}{R_{cl}}\pqty*{n+1}  
% \end{equation}
% which leads to the dispersion relation
% \begin{equation}
%   q_0 = \sqrt{\frac{2 \mu}{3}\pqty*{q^2 - \frac{2 \omega^2}{\mu}}} = \sqrt{\frac{2\mu}{3}} q \pqty*{1 - \frac{\omega^2}{q^2 \mu} + \dots}
% \end{equation}

% * conclusions
%%%%%%%%%%%%%%%%%%%%%%%%%%%%%%%%%%%%%%%%%%%%%%%%%%%%%
\section{Conclusions}
\label{sec:conclusions}

Experiments with ultracold atomic gases invariably impose a trapping
mechanism in order to confine the atoms to a controlled, finite region
of space. As a consequence, measurements of atomic observables carry
an imprint of the trapping potential, denoted by $V$, and conclusions
regarding the homogeneous system—i.e. the cold atomic gas in the
absence of the trap—require a detailed quantitative understanding of
the trapping effects. A common and very-successful method for managing
these effects is the \acl{lda} which
effectively replaces everywhere the chemical potential with a
position-dependent quantity which incorporates the effect of the
trapping potential. In this paper, we have performed a careful study
of the trap dependence of the superfluid \ac{eft} which describes fermions
at unitarity. The sole degree of freedom in this \ac{eft} is the phonon
field, whose fluctuations are strongly modified by V. This particular
system is of tremendous topical interest experimentally,
and is consequently also of great interest to theorists, who are able
to exploit the large degree of symmetry of the system.

At unitarity, the Schr\"odinger-invariant superfluid system has a
single scale, the chemical potential $\mu$, generated by spontaneous
breaking of the particle number, and, consequently, the non-relativistic conformal symmetry. The trapping
potential introduces a new scale, which can be taken either as the
size of the superfluid droplet, $R_{cl}$, or the energy scale
\(\varpi\).  These scales suggest a convenient hierarchy where the
phonon field fluctuations can be treated in three distinct regimes
characterized by the energy $q_0$ where $\varpi \ll q_0\ll \mu$.

The main conclusions of this paper are:

\begin{itemize}

\item 
The low-energy regime corresponds to $\varpi \ll q_0 \ll \sqrt{\varpi \mu}$ and the phonon dynamics is described by the \ac{lo} \ac{eft} and higher-order terms in the \ac{wkb} expansion.

\item
The high-energy regime corresponds to  $\sqrt{\varpi \mu} \ll q_0 \ll \mu$ and the dynamics is described by \ac{lo} \ac{wkb} and higher-order terms in the \ac{eft}.

\item
The low- and high-energy regimes are separated by an intermediate linear zone where $q_0 \approx \sqrt{\varpi \mu}$; this is described by
\ac{lo} \ac{eft} and \ac{lo} \ac{wkb} (see Fig.~\ref{fig:energy-scales}).
  
\item
  Through a study of the phonon field fluctuations and dynamic structure factor in the various regimes, we find that the phonon dispersion relation receives significant trap-induced modifications only in the low-energy regime; the strongest effect appears for a harmonic potential. The corrections are small but calculable in the other regimes.
  
\end{itemize}

In this paper, we have treated in detail the special case of power-law
confining potentials. However, the fundamentals of the analysis persist
for general potentials that may be implemented experimentally. In addition,
the results obtained here, while perhaps currently most relevant to
the study of the \ac{bcs}--\ac{bec} crossover regime in gases of fermionic atoms,
have universal applicability. For instance, it may be the case that
the surfaces, or skins, of neutron-rich nuclei (including neutron stars)
experience squeezing of the three-dimensional superfluid
neutron matter into a quasi-two-dimensional region, as well as other
confining scenarios, which can be conveniently modeled by using anisotropic trapping
potentials~\cite{Kanada_En_yo_2009} and analyzed via the methods introduced here.

It is a straightforward matter to include Schr\"odinger-symmetry
breaking effects due to a finite scattering length in the superfluid
\ac{eft}. This enables a study of the fluctuations across the \ac{bcs}--\ac{bec}
crossover. The introduction of yet another scale generalizes the
the simple \ac{eft} hierarchy treated here in a straightforward manner. As
the low-energy constants that appear in the superfluid \ac{eft} are
somewhat uncertain, we postpone the discussion of symmetry breaking to
a separate paper which will address this issue, as well as
finite-temperature corrections, in the context of the large-N
expansion.

\section*{Acknowledgments}

\begin{small}\sffamily
The authors would like to thank S.~Aoki, A.~Bulgac, J.A.~Carlson, S.~Gandolfi,
H.W.~Hammer, S.~Hellerman, D.B.~Kaplan, M.M.~Forbes, D.R.~Phillips, and D.T.~Son
for useful discussions.  This work was supported by the Swiss National
Science Foundation under grant number 200021\_219267. In addition,
S.R.B is supported by the U.~S.~Department of Energy grant
\textbf{DE-FG02-97ER-41014} (UW Nuclear Theory).  D.O. and
S.R. gratefully acknowledge support from the Yukawa Institute for
Theoretical Physics at Kyoto University, as well as the from the
22nd Simons Physics Summer Workshop: \emph{Future pathways for fundamental physics}, at Stony Brook University,
where some the research for this paper was performed.
\end{small}

% * appendices and stuff

\appendix
\DeclareDerivative{\fdv}{\delta}[style-var=multiple]

\section{Response function in the path integral formalism}
\label{sec:resp-funct-path-integral}

We can compute the response function and the dynamic form factor by using a different formalism, based on the path integral quantization.
By definition, the response function is the two-point function for the charge density
\begin{equation}
  \chi(\mathbf{x}, \mathbf{x}' ; t - t') = \ev{ \rho(\mathbf{x},t) \rho(\mathbf{x}',t')}_T \, ,
\end{equation}
where \(\rho(\mathbf{x},t )\) is the charge density.
Note that we are assuming time-translation invariance but no space-translation invariance so to accommodate a potential.
In the path integral formalism, the two-point function is computed by coupling the system to an external electric potential \(A_0\), which is by construction the current dual to the charge density:
\begin{equation}
  \chi(\mathbf{x}, \mathbf{x}' ; t - t') = \fdv{W[A_0]}{{A_0(\mathbf{x},t)},{A_0(\mathbf{x}',t')}} \, ,
\end{equation}
where \(W[A_0]\) is the generating functional
\begin{equation}
  W[A_0] = - \log*( \int \DD{\theta} e^{-S[\theta, A_0]})  \, ,
\end{equation}
where we assume Euclidean signature.

We are interested in the study of small fluctuations, so we expand \(S[\theta, A_0]\) at quadratic order both in the Goldstone field \(\pi = \theta - \mu t\) and in \(A_0\), to write the action in the form
\begin{equation}
  S[\pi,A_0] = \frac{1}{2} \int \dd{t} \dd{\mathbf{x}} \bqty*{ \pi(\mathbf{x}, t) \Op{\pi\pi} \pi(\mathbf{x},t) + 2 \pi(\mathbf{x},t)  \Op{\pi A} A(\mathbf{x},t) + A_0(\mathbf{x}, t) \Op{AA}  A_0(\mathbf{x},t) }  ,
\end{equation}
where \(\Op{ij}\) are local differential operators.
In the \ac{nlo} \ac{eft}, the operator \(\Op{\pi\pi}\) depends on the three low-energy constants \(c_0\), \(c_1\)
and \(c_2\), while \(\Op{\pi A}\) and \(\Op{AA}\) only depend on \(c_0\) and \(c_1\), since the term proportional to \(c_2\) in the action vanishes on the ground state.

The fluctuations are quadratic, and we can perform the path integral to find
\begin{multline}
  W[A_0] = \frac{1}{2} \log(\det(\Op{\pi\pi}))) - \frac{1}{2} \int \dd{x} \dd{x'} A_0(\mathbf{x})\Op{\pi A}^{\dagger}(x)\Op{\pi A}^{\dagger}(x') D(x; x') A_0(x') \\
  + \frac{1}{2}\int \dd{x} A_0(x) \Op{AA}(x) A_0(x) \, ,
\end{multline}
where \(D(x; x') = \Op{\pi\pi}^{-1}\) is the propagator of the small fluctuations.
The response function is then%
\footnote{%
The second term is the static response function:
\begin{equation}
  \chi(\mathbf{x}, \mathbf{x}') = \int \dd{t} \chi(\mathbf{x}, \mathbf{x}' ; t) = \delta(\mathbf{x}- \mathbf{x}') \Op{AA}(\mathbf{x})  
\end{equation}
and, by what we said above, only depends on \(c_0\) and \(c_1\).
}
\begin{equation}
  \chi(\mathbf{x}, \mathbf{x}'; t-t') =-\Op{\pi A}^{\dagger}(x)\Op{\pi A}^{\dagger}(x') D(x; x') + \Op{AA}^{\dagger}(x) \delta(x - x')  \, .
\end{equation}

%%%%%%
To relate it to an experimentally-measurable quantity, it is convenient to introduce relative and center-of-mass coordinates
  \begin{align}
    \mathbf{r} &= \mathbf{x} - \mathbf{x}', &  \mathbf{R} = \frac{\mathbf{x} + \mathbf{x}'}{2},
  \end{align}
  and Fourier transform with respect to \(\mathbf{r}\) and \(\Delta t=t - t'\):
  \begin{equation}
    \chi(\mathbf{q}, \mathbf{R}; q_0) = \int \dd{\Delta t} \dd{\mathbf{r}} e^{i \mathbf{q} \cdot \mathbf{r} - i q_0 \Delta t} \chi(\mathbf{r}, \mathbf{R}; \Delta t) \, .
  \end{equation}
Due to the fluctuation-dissipation theorem, the imaginary part of the average over the droplet is the dynamic structure factor
  \begin{equation}
    S(q_0, \mathbf{q}) = -\frac{1}{\pi}\int \dd{\mathbf{R}} \Im( \chi(\mathbf{q},\mathbf{R}; q_0)) \, .
  \end{equation}
  This function will be peaked along a curve in the \((q_0,q)\) plane.
  The curve \(q_0 = q_0(q)\) is the \emph{dispersion relation}.
%%%%%%

In the \ac{lo} \ac{eft}, the operators are
\begin{align}
  \Op{\pi\pi} &= - \del_0^2{} + L ,  & \Op{\pi A} &=- \partial_0, & \Op{AA} &= 1,
\end{align}
where \(L\) is the Sturm--Liouville operator.
\begin{equation}
  L \pi = \frac{	\frac{\dd{}}{\dd{r}}\pqty*{r^2 \Lag'[\mu- V] \frac{\dd{}}{\dd{r}}\pi } - \Lag'[\mu- V] \ell(\ell+1) \pi }{r^2 \Lag'' [\mu- V]}.
\end{equation}
Using the general theory outlined in Section~\ref{sec:fluctuation-spectrum}, we can decompose the propagator on a basis of eigenfunctions of \(L\):
\begin{align}
  L \pi_n &= E_n^2 \pi_n ,\\
  D(\mathbf{x}, \mathbf{x}' ; q_0) &= \sum_n \frac{\pi_n(\mathbf{x})^{*} \pi_n(\mathbf{x})}{q_0^2 - E_n^2} ,
\end{align}
and recover the same expression for \(S(\mathbf{q}, q_0)\) as in Eq.~\eqref{eq:S-Fermi-golden}.
To go beyond \ac{lo}, it becomes cumbersome to decompose the propagator on the basis of eigenfunctions, but it is convenient to use an alternative approach, as done in~\cite{Hellerman:2023myh}.
We leave this problem for future investigations.

\setstretch{1}

% \newpage
%\printbibliography{}
\small\sffamily

\bibliography{bibi,non-arxiv}{}
\bibliographystyle{JHEP} % We choose the "plain" reference style

\end{document}